\documentclass[12pt,tightenlines,eqsecnum,floats,aps,amsmath,amssymb,tightenlines,nofootinbib,prd,shownopacs,notitlepage]{revtex4}

\usepackage{graphicx,verbatim}
\usepackage{amsmath}
\usepackage{amsfonts}
\usepackage{amssymb}
\usepackage[colorinlistoftodos]{todonotes}
\usepackage{epstopdf}

\newcommand{\pb}[1]{\hbox{\lower0.5ex\hbox{${}_{\leftarrow}$}}\kern-1.9ex{#1}}
\def\chi{\theta}
\def\hchi{\hat\theta}
\def\hTheta{\hat\Theta}

\def\be{\begin{equation}}
\def\ee{\end{equation}}
\def\ba{\begin{eqnarray}}
\def\ea{\end{eqnarray}}

\def\t{\tilde}
\def\h{\hat}

\def\ub{\underbar}
\def\ul{\underline}
\def\f{\frac}
\def\ub{\underbar}
\def\x{\underline{\underline{\eta}}}

\def\rmd{\mathrm{d}}
\def\S{\mathcal{S}}
\def\F{\mathcal{F}}
\def\A{\mathcal{A}}
\def\H{\mathcal{H}}
\def\W{\mathcal{W}}
\def\g{\mathfrak{g}}
\def\E{\mathcal{E}}
\def\e{\mathfrak{e}}

\def\eal{{\mathring{e}_{(\beta)}}}

\def\hphi{\hat{\phi}}
\def\hPhi{\hat{\Phi}}
\def\stress{\langle \hat{T}_{ab}(x) \rangle_{\rm ren}}
\def\phisq{\langle \hat{\phi}^2(x)\rangle_{\rm ren}}

\def\Mo{\mathring{M}}
\def\no{\mathring{n}}
\def\go{\mathring{{g}}}
\def\phio{{\phi}^\circ}
\def\Do{\mathring{D}}
\def\hphio{{\hat{\phi}^\circ}}
\def\biphi{\langle \h\phi(x)\,\h\phi(\xp)\rangle}
\def\fo{\mathring{f}}
\def\Deltao{\mathring{\Delta}}
\def\Boxo{\mathring{\Box}}
\def\Fo{\mathring{F}}
\def\Foub{\mathring{\underbar{F}}}
\def\Ao{\mathcal{A}^\circ}
\def\Aoub{\underline{\mathcal{A}}^{{}^\circ}}
\def\Omegao{\mathring\Omega}

\def\eoub{\mathring{\underbar{e}}}
\def\hPhi{\h\Phi}
\def\hPhio{{\h\Phi}^\circ}
\def\hchi{\hat\chi}
\def\bichi{\langle \h\theta(x)\,\h\theta(\xp)\rangle}
\def\etao{\eta_{\circ}}
\def\pso{\mathring{\Gamma}_{\rm cov}}
\def\psoub{\mathring{\underline{\Gamma}}_{\,\rm cov}}
\def\Jo{\mathring{J}}
\def\Jub{\underbar{J}}
\def\Joub{\mathring{\!\underbar{J}}}
\def\Omegaoub{\mathring{\underline{\Omega}}}
\def\Houb{\,\,\mathring{\!\!\underline{\mathcal H}}}
\def\Fockoub{\mathring{\underline{\mathcal F}}}
\def\Focko{\mathring{\mathcal{F}}}
\def\Vo{\mathring{V}}

\def\ps{\Gamma_{{\rm Cov}}}

\def\vx{\vec{x}}
\def\vk{\vec{k}}
\def\vr{\vec{r}}
\def\vxp{\vec{x}^{\,\prime}}
\def\etap{\eta^{\,\prime}}

\def\xp{x^{\,\prime}}

\def\Ak{\hat{A}(\vec{k})}

\def\Admk{{\hat{A}^{\dagger}(-\vec{k})}}

\begin{document}

\title{Probing the Big Bang with quantum fields}

\author{Abhay Ashtekar${}^{1}\,$}
\email{ashtekar.gravity@gmail.com} \author{Tommaso De Lorenzo${}^{1}\,$}
\email{tommasodelorenzo@yahoo.it}
\author{Marc Schneider ${}^{1,2}\,$}
\email{marc.schneider13@googlemail.com}
\affiliation{${}^{1}$ Institute for Gravitation and the Cosmos \& Physics Department,\\ The Pennsylvania State University, University Park, PA 16802 U.S.A}
\affiliation{${}^{2}$ Albert Einstein Institute for Gravitational Physics\\ Am M\"uhlenberg 1, \, 14476 Potsdam, Germany} 

\begin{abstract}

By carrying out a systematic investigation of linear, test quantum fields $\h\phi(x)$ in cosmological space-times, we show that $\h\phi(x)$ remain well-defined across the big bang as \emph{operator valued distributions} in a large class of Friedmann, Lema\^itre, Robertson, Walker space-times, including radiation and dust filled universes. In particular, the expectation values $\biphi$ are well-defined bi-distributions in the extended space-time in spite of the big bang singularity. Interestingly, correlations between fields evaluated at spatially and temporally separated points exhibit an asymmetry that is reminiscent of the Belinskii, Khalatnikov, Lifshitz behavior. The renormalized products of fields $\phisq$ and $\stress$ also remain well-defined as distributions. Conformal coupling is not necessary for these considerations to hold. Thus, when probed with observables associated with quantum fields, the big bang (and the big crunch) singularities are quite harmless. 
\end{abstract}

\maketitle

\section{Introduction}
\label{s1}

The celebrated singularity theorems of general relativity are based on the mathematical notion of geodesic incompleteness, which translates physically to the statement that trajectories of test particles come to an abrupt end. However, it has long been argued that the effect of space-time singularities would be softened if one uses physically more appropriate probes such as test fields. For example, the naked singularities of certain static, dilatonic, extremal black holes have been shown to be tame when probed with \emph{quantum} test particles \cite{hm}:  
their dynamics continues to be well-defined even though these space-times are of course geodesically incomplete. Similarly, it has been shown that the evolution of classical test fields can be meaningfully defined in presence of the naked singularity even in the super extremal Reissner-Nordstr\"om solutions \cite{stz}. It has also been argued that evolution of \emph{quantum fields} does not break down at the Schwarzschild singularity \cite{hs}, albeit the reasoning relies on formal arguments that does not do full justice to the difficult quantum field theoretic issues associated with an infinite number of degrees of freedom.

In this paper we turn to the \emph{time-dependent} metrics with space-like 
singularities \emph{that result dynamically}. In the main body of the paper we will focus on the simplest cases provided by Friedmann, Lema\^itre, Robertson, Walker (FLRW) cosmologies. The FLRW space-times are of course geodesically incomplete and the tidal force on test particles diverges because of the curvature singularity. Turning to classical test fields one finds that smooth initial data generically yield solutions that diverge at the big bang. In this sense the singularity is equally bad for classical fields. What would happen if one uses test \emph{quantum probes} instead? Because the geometry is now time dependent, in contrast to the static situation considered, e.g., in \cite{hm} now we have to use \emph{quantum fields} rather than quantum particles as probes. Are these singularities as disastrous for quantum fields as they are for their classical analogs? Or, do they appear tame? Our goal is to present a systematic investigation of this issue. Although we focus on certain FLRW space-times in order to perform explicit analytical calculations in the main body of the paper, as we indicate in the Discussion section and Appendix \ref{a2}, the situation would be qualitatively similar in a much larger class.

In this analysis, the key conceptual and mathematical difference between a classical field $\phi(x)$ and its quantum analog $\h\phi(x)$ turns out to be the following. In physically interesting cases one expects $\phi(x)$ to be a suitably smooth function that obeys its field equation in the corresponding smooth sense. The field $\phi(x)$ senses the singularity via a blow up of its numerical value there. On the other hand, already in Minkowski space, the quantum field $\hphi(x)$ is an operator valued \emph{distribution} (OVD) \cite{asw}. Therefore, we are led to ask whether it continues to be well-defined as an OVD across the big-bang in FLRW space-times. 
Do the smeared operators $\hphi(f) := \int \rmd^4V \hphi(x) f(x)$ remain well-defined and continue to satisfy the field equation $\int \rmd^4 V\,\hphi(x)\, (\Box -m^2) f(x) =0$ even when the support of the test field $f(x)$ includes the singularity? Here $\rmd^4 V$ is, as usual, the space-time volume element determined by the FLRW metric.  Now, the volume element shrinks at the big bang (as well as at the Schwarzschild singularity). Therefore, it is `easier' for $\h\phi(x)$ (as well as the classical solutions $\phi(x)$) to be a well-defined distribution. A simple analogy is provided by $h(\vec{r}) := 1/r$ on $\mathbb{R}^3$: it is singular as a function but well-defined as a distribution (satisfying the well known equation $\vec\nabla^2\, h(\vec{r}) = - 4\pi\, \delta(\vec{r})$, again in the distributional sense). We will find that, as operator-valued distributions, the space-time quantum fields $\hphi(x)$ remain perfectly well-defined across cosmological singularities (and satisfy the field equation in a distributional sense). In particular, the expectation values $\langle \h\phi(x) \h \phi(\xp)\rangle$ are well-defined bi-distributions (in the naturally extended space-time) in spite of the curvature singularity at the big bang. Thus the usual nomenclature `field operators' and `2-point functions' can be quite misleading in the investigation of the effect of singularities; their distributional character makes a crucial difference. In addition, we will find that, as one approaches the singularity, there is an interesting effect: correlations between fields at spatially separated points behave differently from those between temporally separated points, an asymmetry that is reminiscent of the Belinskii, Khalatnikov, Lifshitz behavior \cite{bkl,berger}. 

Next one can ask for the behavior of the renormalized expectation values of \emph{products} of fields. The most commonly used are $\phisq$ and $\stress$. It turns out that they also remain well defined as distributions across the big bang (and the big crunch). Note that $\hphi(f)$ are `dimension 1' operators (that do not directly `sense' the Riemann curvature), $\phisq$ are `dimension 2' observables (that are sensitive only to the scalar curvature), while $\stress$ are dimension 4 observables  (sensitive to products of curvature tensors). While all these observables continue to be well-defined in the distributional sense, we will find that the lower the physical dimension, `tamer' is the detailed distributional character in conformal time. 

The paper is organized as follows. In section \ref{s2} we recall the elements of the quantum field theory in cosmological space-times that we will need in our detailed analysis. In section \ref{s3} we probe the big bang singularity of the FLRW space-times that are sourced by radiation using a quantum field $\h\phi(x)$. These space-times are rather special in that the Ricci scalar vanishes there, which simplifies the analysis considerably. As a prelude to considering more general FLRW space-times, in section \ref{s4} we make a detour to study the behavior $\h\phi(x)$ in Minkowski space but in presence of certain time-dependent potentials which diverge as $1/\eta^2$ at time $\eta=0$. These results are then readily transferred to the dust filled FLRW universe in section \ref{s5}, and more general FLRW space-times in Appendix \ref{a2}. In section \ref{s6} we collect the main results and put them in a broader perspective. In Appendix \ref{a1} we collect some results \cite{schwartz,gs,hormander,loja} on tempered distributions (\'a la Schwartz) that clarify the origin of ambiguities that often result in regularization procedures. In Appendix \ref{a2} we briefly discuss various generalizations of the main results.

Finally, our conventions are as follows. We use the signature $-,+,+,+$ and define the curvature tensors via $ 2 \nabla_{[a} \nabla_{b]} K_c = R_{abc}{}^d K_d$; \, $R_{ac} = R_{abc}{}^b$ and $R = g^{ab} R_{ab}$. In the main body of the paper we will restrict ourselves to scalar fields satisfying the Klein-Gordon equation and only briefly discuss more general examples in Appendix \ref{a2}. However, even there we will only consider bosonic fields. We believe that tameness of the singularity will persist also for fermionic fields although certain issues will require greater attention (e.g. the relative orientation of frame fields on the two sides of the big bang).

\section{Preliminaries}
\label{s2}

In this section we collect the results that provide the conceptual framework for the subsequent sections. This discussion will also serve to fix the notation. In the first part, we discuss the algebra $\A$ of operators for linear fields in globally hyperbolic space-times. This algebra can be constructed using just the classically available structure. In the second part we recall the new mathematical input needed to construct quasi-free representations of $\A$, and in the third part we summarize how this structure is generally specified for quantum fields on FLRW space-times.

\subsection{The operator Algebra}
\label{s2.1}

Let $(M, g_{ab})$ be a globally hyperbolic space-time. For definiteness, we will focus on test quantum scalar fields $\hphi (x)$ on this space-time satisfying $(\Box - m^2) \hphi (x) =0$, although our considerations can be easily generalized to include the scalar fields that satisfy the conformally invariant equation, as well as Maxwell and linearized gravitational fields. $\hphi (x)$ is an  OVD on $M$ and the field operators $\hphi(f)$ are obtained by smearing $\hphi (x)$ with suitable test fields $f(x)$: $\hphi(f) = \int_M \rmd^4 V\, \hphi(x) f(x)$, where $\rmd^4 V$ is the volume element on $M$ induced by $g_{ab}$. (The precise space of test functions will be specified in section \ref{s2.3}). The algebra $\A$ of interest is the (abstractly defined) free $\star$-algebra generated by $\hphi(f)$ satisfying the self-adjointness relation $\hphi^\star(x) = \h\phi(x)$,
the commutation relations
\be \label{comm1} [\hphi(x),\, \hphi(\xp) ] = i{\hbar}\, \Delta (x,\,\xp)\, \h{I},\ee
or, equivalently
\be \label{comm2}
    [\hphi(f_1),\, \hphi(f_2) ] = \,{i\hbar}\int_M\! {\rmd}^4 V f_1(x)
    \int_M {\rmd}^4 V^\prime \,\Delta (x,\,\xp)\,  f_2(\xp)\,\,\h{I} , \ee
and the field equations
\be \label{fe}  \hphi ((\Box-m^2)f) := \int_M \!\rmd^4V\, \hphi(x)\, (\Box - m^2) f(x) =0 \ee 
for all test functions $f_1(x), f_2(x)$ and $f(x)$. The distribution $\Delta (x,\, \xp)$ on the right side of (\ref{comm1}) is the difference between the advanced and retarded Green's functions, $\Delta (x,\, \xp) := (G_{\rm Ad} \, - G_{\rm Ret})(x,\,\xp)$  \cite{lichne}. Thus, the algebra $\A$ can be constructed using structures that are already available on $(M, g_{ab})$. 

Note that (\ref{fe}) implies that the map $f \to \hphi(f)$ has a huge kernel since $\hphi(f) =0$ if $f$ is of the form $(\Box-m^2)g(x)$ for any test field $g(x)$. It is often convenient to remove this kernel as follows. Since $\Delta (x,\,\xp)$ satisfies the field equation in both arguments, it follows that 
\be \label{sol1} F(x) := \int_M \rmd^4 V^\prime  \Delta(x,\, \xp) f(\xp) \ee
is a solution to the field equations. This correspondence $f(x) \to F(x)$ from the space of test functions to solutions of the field equation has the property that if we replace the test field $f(x)$ with $\t{f} = f + (\Box -m^2)g$, for any test field $g$ we have $\t{f} \to F(x)$. Thus, at the classical level, the map (\ref{sol1}) from test fields to solutions removes the kernel.

Can we then remove the redundancy also in the quantum theory? The answer is in the affirmative \cite{am}. Recall that, restrictions of the commutator Green's function $\Delta(x,\,\xp)$ and its time derivative to a Cauchy surface $t=t_0$ satisfy: (i) $\Delta(x,\,\xp)\mid_{t=t^\prime =t_0}\, =\,0$, and (ii) $n^a \nabla_a\, \Delta(x,\,\xp)\mid_{t=t^\prime=t_0}\, = \,\delta^3 (\vx, \vxp)$, where $n^a$ is the unit normal to the Cauchy surface. (These properties of $\Delta(x,\,\xp)$ and (\ref{comm1}) imply that $\hphi(\vx)$ and its conjugate momentum $\h\pi(\vx)$ satisfy the standard canonical commutation relations). Next, note that the symplectic inner product can be evaluated on any Cauchy surface. Therefore, for any given $x'$ let us evaluate the symplectic product between a classical solution $\phi(x)$ and $\Delta(x,\,\xp)$ at time $t=t^\prime$ and obtain
\be \label{Omega} \Omega(\phi(x),\, \Delta(x,\,\xp)) := \int_{t=t^\prime}\!\! \rmd^3 V \big[\phi(x) (n^a\nabla_a \,\Delta(x,\,\xp))-  (n^a \nabla_a \phi(x)) \Delta(x,\,\xp)\big] \, =\, \phi(\xp)\ee
Replacing $\phi(x)$ with the  OVD $\hphi(x)$ and smearing both sides of the final equality with a test function $f(\xp)$ we obtain:
\be  \Omega(\hphi(x),\, F(x)) = \hphi(f) \ee
for all test functions $f$. Finally, let us set $\hPhi(F) := \Omega(\h\phi(x),\, F(x))$. Then, the commutation relations (\ref{comm2}) are replaced by
\be [\hPhi(F_1),\, \hPhi(F_2)] =  i\hbar\, \Omega(F_1,\, F_2)\, \h{I} \ee
bringing to forefront the relation between quantum commutators and the symplectic structure on the classical phase space. Furthermore, since test fields $f(x)$ and $f(x) + (\Box - m^2)g(x)$ define the same solution $F(x)$ for all test fields $g(x)$, the redundancy in the label is removed: $\hPhi(F) =0$ if and only if $F(x) =0$. To summarize, we can generate the algebra $\A$ either using the field operators $\hphi(f)$ smeared with test fields $f$ (for which the map $f \to \hphi(f)$ has a kernel) or with operators $\hPhi(F)$ associated with solutions $F$ (for which the map $F \to \hPhi(F)$ is faithful). We will find both versions of field operators are useful in probing the big bang singularity with quantum fields.

On Hilbert spaces of interest to our analysis, the operators $\hphi(f)$ and $\hPhi(F)$ are represented by unbounded self-adjoint operators. We note that it is often convenient to work with their exponential versions  $\h{W}(F) := \exp \f{i}{\hbar} \hPhi(F)$ which are represented by unitary operators. The Weyl algebra $\W$ generated by $\h{W}(F)$ is particularly convenient because the vector space of their linear combinations is closed under the product:
\be \label{prod} \h{W}(F_1) \h{W}(F_2) = e^{-\f{i}{2\hbar} \Omega(F_1, F_2)}\, \h{W}(F_1+F_2)\, .\ee
Thus $\W$ is spanned just by linear combinations of $\h{W}(F)$.

\subsection{Quasi-free Representations}
\label{s2.2}

While the algebras $\A$ and $\W$ can be constructed using structures that are naturally available classically on general globally hyperbolic space-times \cite{lichne,am}, one needs a new mathematical structure to find their representations on Hilbert spaces. The most-commonly used representations are the `quasi-free ones' in which the Hilbert space has the structure of a Fock space and the field operators $\hphi(f)$ and $\hPhi(F)$ are represented by sums of creation and annihilation operators associated with one particle states \cite{shale,segal}. In this case, the necessary new element is a complex structure $J$ on the space of classical solutions $F$. Denote by $\ps$ the covariant phase space of suitably regular solutions $F(x)$ to the classical field equation $(\Box - m^2) F(x) =0$. (The regularity conditions will be spelled out in section \ref{s2.3}.) Then a complex structure $J$ is a linear map on $\Gamma$ satisfying $J^2 = -1$. $\ps$ is already endowed with a symplectic structure $\Omega$ (of (\ref{Omega})). The complex structure $J$ is required to be compatible with $\Omega$ in the sense that\,\, $\g (F_1,\, F_2) := \Omega( F_1,\, J F_2)$\, is a positive definite metric on $\ps$. Then the triplet $(\Omega, J, \g)$ endows $\ps$ with the structure of a K\"ahler space. In particular, then,
\be \label{ip1} \langle F_1,\,  F_2\rangle = \f{1}{2\hbar}\, \big(\g(F_1,\, F_2) + i \Omega (F_1,\, F_2) \big) \ee
is an Hermitian inner product on the complex vector space  $(\ps, J)$. The Cauchy completion $\H$ of $(\ps, J, \langle \cdot\, ,\, \cdot\rangle)$ provides us with a 1-particle Hilbert space. The symmetric Fock space $\F$ generated by $\H$ is the Hilbert space that carries a quasi-free representation of $\A$ (and $\W$). Specifically, with each 1-particle state $F$ (now regarded as an element of $\H$), there is an annihilation operator $\h{A}(F)$ and a creation operator $\h{A}^\dag (F)$ on $\F$, satisfying the commutation relation\, 
\be \label{comm4} [ \h{A}(F_1), \, \h{A}^\dag(F_2)] = \langle F_1,\, F_2\rangle\, \h{I}\, .\ee
The abstract field operators $\hphi(f) = \hPhi(F)$ are represented by concrete operators on $\F$ as follows:
\be \hPhi(F) = \hbar \big(\h{A}(F) + \h{A}^\dag (F) \big)  \ee
For details see \cite{am,am-grg}.

In terms of the Weyl algebra $\W$, the Gel'fand-Naimark-Segal (GNS)  \cite{gns1,gns2} construction provides a direct and more elegant avenue to introduce this representation. Given a positive linear function on a $\star$-algebra --in physical terms, an expectation value function $\E$-- the construction provides an explicit, step by step procedure to build a Hilbert space and represent elements of the algebra as concrete operators on it such that all algebraic relations are preserved. In the case of the Weyl algebra $\W$ then the task is to provide a linear, complex valued function $\E$ on $\W$  such that  $\E(A^\star A) \ge 0$ for all $A \in \W$. Because of the property (\ref{prod}) of the Weyl operators, the task reduces to finding a complex-valued function $\mathfrak{e}$ on $\ps$ such that  $\E(W(F)) := \mathfrak{e}(F)$ is a positive linear function on $\W$. Given a complex structure $J$ on $\ps$, the required $\e$ is determined by the Hermitian product it defines on $\ps$: 
\be \label{plf1} \e (F) = \, e^{ -\f{1}{2} \langle F, F\rangle}\, \equiv \, e^{ -\f{1}{4\hbar} \Omega(F, JF)}\, .\ee

\subsection{FLRW Space-times}
\label{s2.3}

Let us now specialize to spatially flat FLRW space-times $(M, g_{ab})$ with spatial topology $\mathbb{R}^3$. Since the space-time metric $g_{ab}$ is conformally flat, we have:
\be  \label{metric} {g}_{ab}\rmd x^{a} \rmd x^{b} =  a^{2}(\eta)\, \go_{ab}\rmd x^{a} \rmd x^{b} \equiv
a^{2}(\eta) \big(-\rmd \eta^{2} + \rmd \vx^{2}\big), \ee
so that $\eta$ is the conformal time coordinate, related to proper time $t$ via $a(\eta) \rmd\eta = \rmd t$, and $(\eta,\vx)$ are the Cartesian coordinates of the  Minkowski metric $\go_{ab}$. In the main text we will restrict ourselves to the massless scalar field satisfying\, $\Box \phi(\vx, \eta) =0$\, on $(M, g_{ab})$, and in Appendix \ref{a2} we briefly discuss more general scalar fields as well as higher spin fields. The overall conceptual picture is the same in these more general cases.

Using the form (\ref{metric}) of $g_{ab}$, the Klein Gordon equation can be readily cast in the form
\be
\phi^{\prime\prime} - \Do^2 \phi + 2 \f{a'}{a}\, \phi' = 0
\label{EOM}
\ee
where primes refers to derivative with respect to conformal time $\eta$, and $\Do^2$ is the spatial Laplacian defined by $\go_{ab}$. By exploiting spatial flatness, we can carry out a Fourier decomposition
\be \label{FT}
\phi(\vx,\eta) = \f{1}{(2\pi)^3} \int \!\rmd^3 k\, \, \phi(\vk,\,\eta) \, e^{i\, \vk \cdot \vx}\, .
\ee 
Because $\phi(\vx,\eta)$ is real, the Fourier transforms are subject to the `reality condition'\, ${\phi}^\star(\vk,\, \eta) = \phi (-\vk,\, \eta)$. The equation of motion (\ref{EOM}) implies that $\phi(\vk,\,\eta)$ satisfies a second order ordinary differential equation in $\eta$ for each $k$. Let us introduce a  basis $e(k,\,\eta)$ of solutions to this equation:
\be  \label{EOM2} e^{\prime\prime} (k,\,\eta) + 2 \f{a^{\prime}(\eta)}{a(\eta)}\, 
e^\prime(k,\,\eta) + k^{2} e(k,\,\eta) =0\, ,\ee
satisfying the normalization condition
\be \label{normalization} e(k,\,\eta)\,{e}^{\star\,\prime} (k,\,\eta)- e^{\prime} (k,\,\eta)\,{e}^\star(k,\,\eta)=\frac{i}{a^2(\eta)}\, ,\ee
for each $k = |\vk|$. Expanding $\phi(\vk,\, \eta)$ in this basis, we obtain 
\be\label{expfield}
\phi(\vx,\eta) = \int\, \f{\rmd^3 k}{(2\pi)^3} \,\big(z(\vk)\, e(k,\,\eta) + {z}^\star(-\vk) \, e^\star(k,\, \eta) \big) \; e^{i\, \vk \cdot \vx}\, \ee
for some complex-valued functions $z(\vk)$. We will assume that $z(\vk)$ belong to the \emph{Schwartz space}\, $\t\S$\,  \cite{schwartz} consisting of smooth functions in the (3-dimensional) momentum space that, together with all their derivatives, fall-off faster than any polynomial as $|\vk| \to \infty$. Given a basis $\{e(k, \eta)\}$, then, we obtain a convenient space of solutions to the Klein-Gordon equations via (\ref{expfield}). This will be our covariant phase space $\ps$. Note that, in contrast to $\phi(\vk)$, \emph{there is no relation between $z(\vk)$ and $z(-\vk)$}. Therefore $z(\vk)$ serve as complex (Bargmann) coordinates  \cite{bargmann1,bargmann2} on the real phase space $\ps$. 

The set of solutions\, $e(k, \, \eta)\, e^{i\vk\cdot\vx}$\, provides an orthonormal `positive frequency basis' in the 1-particle Hilbert space of states $\H$, analogous to the positive-frequency basis $\f{e^{-ik\eta}}{\sqrt{2k}} e^{i\vk\cdot\vx}$\, in Minkowski space. Therefore, it also provides us with a natural complex structure $J$ on $\ps$:
\be \label{J1}
J\,\phi(\vx,\eta) =  \int\, \f{\rmd^3 k}{(2\pi)^3}\, \big(i\, z(\vk)\, e (k,\,\eta) -i\, {z}^\star(-\vk) \, e^\star(k,\,\eta) \big) \; e^{i\, \vk \cdot \vx}\, .
\ee
Thus, the action of $J$ on $\phi(x)$ multiplies the coefficients $z(\vk)$ of the positive frequency basis functions $e(k, \eta)$ by $i$ and the coefficients ${z}^\star(\vk)$ of the `negative frequency basis functions' by $-i$, so that $J\, \phi(\vx,\eta)$ is again a real solution. We can define a new `positive frequency basis'  $\t{e}(k,\, \eta)$ by taking linear combinations of $e(k,\, \eta)$ with appropriately normalized coefficients; the complex structure defined by $\t{e}(k, \eta)$ is again $J$. Thus, the invariant content in the choice of a positive frequency basis is captured by the complex structure. 

The normalization condition (\ref{normalization}) implies that $J$ is compatible with the symplectic structure $\Omega$ on $\ps$. One can readily verify that the Hermitian inner product of Eq. (\ref{ip1}) takes the form
\be \label{ip2}
\langle \phi_1(\vx,\eta),\, \phi_2(\vx,\eta)\rangle = \f{1}{\hbar}\, \int\f{\rmd^3 k}{(2\pi)^3}\; {z}^{\star}_{1} (\vk) \; z_{2} (\vk)\, .
\ee
The 1-particle Hilbert space $\H$ is the Cauchy completion of $\ps$ with this inner product. Thus, for a generic solution $\phi(\vx, \eta)$ of (\ref{expfield}) to belong to $\H$,\, $z(\vk)$ just has to be in $L^2(\mathbb{R}^3)$. In the Fock representation determined by the complex structure $J$ of (\ref{J1}), the  OVD $\hphi(x)$ is represented as
\be \hphi(\vx,\eta) = \int\, \f{\rmd^3 k}{(2\pi)^3}\, \big(\h{A}(\vk)\, e(k,\,\eta) +\, \h{A}^\dag(-\vk) \, e^\star(k,\,\eta) \big) \; e^{i\, \vk \cdot \vx}\, \ee
where the annihilation and creation operators satisfy the commutation relations:
\be \label{comm3}[\h{A}(\vk)\, ,\, \h{A}^\dag (\vk^\prime)]\, =\, (2\pi)^3\, \hbar \,\delta(\vk, \vk^\prime). \ee
Field operators $\hphi(f)$ will be obtained by smearing $\hphi(\vx,\eta)$ with test fields $f$ that are in the Schwartz space $\S$ on $\Mo$, i.e. that are $C^\infty$ w.r.t. $(\vx, \eta)$ and, together with all their derivatives, fall off faster than any polynomial as these Cartesian coordinates go to infinity. Thus, our distributions will be the (operator-valued and ordinary) tempered distributions \cite{schwartz}. \\

\emph{Remarks:} 

1. Literature on quantum field theory in Minkowski space-time generally uses tempered distributions because the Schwartz space $\S$ is stable under Fourier transform, i.e., the Fourier transform $\t{f}(k)$ of a function $f(x) \in \S$ is again in the Schwartz space. However, the construction of $\S$ is tied to the use of Cartesian coordinates which are not available in general curved space-times. Therefore, in quantum field theory in curved space-times, one works with more general distributions, smeared with test functions that belong to $C_0^\infty (M)$, the space of $C^\infty$ functions of compact support in space-time. We chose to work with tempered distributions because in quantum field theory on FLRW space-times, one does have geometrically defined `Cartesian' coordinates $(\eta, \vec{x})$ and it is convenient to go back and forth between fields on spatially homogeneous, isotropic 3-manifolds and their Fourier transforms. Since $C_0^\infty(\mathbb{R}^4) \subset \S$, the action of our tempered distributions is, in particular, well defined on $C_0^\infty(\mathbb{R}^4)$. \smallskip

2. As noted above, the components $\{ z (\vk) \} $ of $\phi(\vx,\eta)$ in the orthonormal basis $e (k, \eta)$ provides a convenient coordinate system on $\ps$.  As a side remark, we note that Eq. (\ref{ip1}) and (\ref{ip2}) imply that these coordinates are well-adapted to the symplectic structure: their Poisson brackets have the form
\be \label{pb1} 
\{ z(\vk),\, {z}^\star(\vk^\prime)\} = -i\,(2\pi)^3\,\delta(\vk,\,\vk^\prime)\, \qquad {\rm and} \qquad \{ z(\vk),\,  z(\vk^\prime)\} =0\, , \ee 
mirroring the commutation relations (\ref{comm3}) between the creation and annihilation operators.\\ 

In sections \ref{s3} and \ref{s4} we will show that, in the models we discuss in detail, the phase space $\ps$ admits a natural complex structure. In sections \ref{s3} - \ref{s5} we will explore in detail properties of the resulting quasi-free, Fock representations across the big bang. To analyze this issue, we will use the manifold $\Mo$ underlying the full Minkowski space-time, so that each of the Cartesian coordinates $(\vx, \eta)$ of the Minkowski metric $\go_{ab}$ takes values on the \emph{full} real line $\mathbb{R}$. $\Mo$ is an extension of the manifold $M$ underlying the FLRW solution $g_{ab}$ since $\eta$ takes only positive values on $M$, $\eta=0$ being the big bang. Since we are interested in the big bang, we will assume that the scale factor has the form $a(\eta) = a_{\beta} \eta^{\beta}$\, with $\beta > 0$ and extend $g_{ab}$ to all of $\Mo$ using $g_{ab} = a^2(|\eta|) \go_{ab}$. Thus $g_{ab}$ will be a well-defined (at least $C^0$) tensor field on all of $\Mo$,\, but degenerate as a metric at $\eta=0$ since $a(\eta)$ vanishes there. The question is: Do various observables associated with the quantum field $\hphi(x)$ remain well-defined  across $\eta=0$?

\section{The Radiation filled universe}
\label{s3}

In this section, we focus on the FLRW space-time sourced only by a radiation field. In this case, $a(\eta) = a_1 \eta$, whence the scalar curvature vanishes: $R(\eta) = 6a^{\prime\prime}(\eta)/a^3(\eta) =0$. Therefore we can rewrite the massless Klein-Gordon equation in a conformally covariant form  $(\Box - R/6)\,\phi(\vx,\eta) =0$. Since the metric $g_{ab}$ is conformally related to the Minkowski metric $\go_{ab}$ of (\ref{metric}), the analysis simplifies greatly, enabling us to discuss the regularity of the quantum field $\hphi (x)$ across the big bang in a setting that is devoid of non-essential technical complications.

Note that if one were to consider a Maxwell field, the field strength $F_{ab}$ would satisfy source-free equations with respect to $g_{ab}$ if and only if it satisfies them with respect to $\go_{ab}$. This conformal \emph{invariance} implies that regular solutions to the Maxwell equation in Minkowski space remain regular also across the big bang in both classical and quantum field theory. By contrast, the equation $(\Box - R/6)\,\phi(\vx,\eta) =0$ is only conformally \emph{covariant}: If $\phio(x)$ satisfies this equation with respect to the Minkowski metric $\go_{ab}$, then it is $\phi(x) = a^{-1}(\eta) \phio(x)$ that satisfies it with respect to the radiation filled FLRW metric $g_{ab}$. (Recall that symbols with a zero above them (or carrying a superscript $\circ$) refer to the Minkowski space $(\Mo, \go_{ab})$.) Since $a(\eta) =0$ at the big bang, generic classical solutions $\phi(x)$ diverge there. Therefore, in contrast to the Maxwell case, the issue of whether the quantum field $\hphi(x)$ and observables associated with it are regular across the big bang is non-trivial even though the scalar curvature $R$ vanishes.

Our discussion is divided into two parts. In the first we discuss the regularity of $\h\phi (x)$ as an  OVD. In the second we investigate properties of various expectation values -- the 2-point bi-distribution $\langle \hphi(x) \hphi(\xp) \rangle$ and the renormalized products of operator-valued distributions $\phisq$ and  $\stress$.

\subsection{Field operators and the Big Bang}
\label{s3.1}

Let us begin with the massless Klein-Gordon field $\hphio (x)$ on the full Minkowski space-time $(\Mo, \go_{ab})$. The $\star$-algebra $\Ao$ is generated by the smeared operators $\hphio (\fo)$\, (or, $\hPhio (\Fo) = \Omegao (\hphio,\, \Fo)$)\, satisfying the 
field equation
\be \label{feo} \int\! \rmd^4x\, \hphio(x)\, \Boxo f(x) =0 \quad \hbox{\rm for all test fields $f$ in the Schwarz space $\S$}, \ee
and the commutation relations
\be [\hphio (x),\, \hphio(\xp)] = i\hbar\, \Deltao(x,\, \xp)\, \h{I} \quad {\rm or}\quad [\hPhio(\Fo),\, \hPhio(\Fo^{\,\prime})] = i\hbar\, \Omegao (\Fo,\, \Fo^{\,\prime})\, \h{I}. \ee 
As explained above, the Schwartz $\S$ associated with $\Mo$ will consist of smooth functions $f$ that, together with all their derivatives, fall-off faster than any polynomial as $x \equiv (\vx, \eta)$ go to infinity.

Conformal covariance considerations lead us to introduce a candidate operator-valued distribution $\hphi(x)= \hphio(x)/a(\eta)$ on the extended FLRW space-time $(\Mo, \, g_{ab}= a^2(\eta) \go_{ab})$. The question is whether it is well-defined and has the desired properties even though $a(\eta)$ vanishes at $\eta=0$. Given a test field $f(x)\in \S$ we have:
\be \label{ovd1} \int_{\Mo}\! \rmd^4 V\, \hphi(x) f(x)\, =\,  \int_{\Mo}\! \rmd^4 x \,\,\hphio(x)\, (a^3(\eta) f(x))\, \equiv \int_{\Mo}\! \rmd^4 x\,\, \hphio(x) \fo(x) \ee
where we have set $\fo(x) = a^3(\eta) f(x)$. Since $a^3(\eta) = a_1^3 \eta^3$, it follows that $\fo$ is also smooth and, together with all its derivatives, falls off  faster than any polynomial at infinity; i.e., $\fo(x) \in \S$. (Put differently, $a^3(\eta) \hphio(x)$ is a well-defined OVD because it is the product of a $C^\infty$ function $a^3(\eta)$ of $\eta$ that grows only polynomially at infinity and an OVD $\hphio(x)$.) Thus the right side of (\ref{ovd1}) belongs to $\Ao$, whence our $\hphi (x)$ is a well-defined  OVD on $(\Mo, g_{ab})$. (Note that vanishing of the volume element\, $\rmd^4 V = a^4(\eta) \rmd^4 x$\, at $\eta=0$ plays a crucial role in this and subsequent arguments.) Next, let us consider the field equation. For any test field $f(x) \in \S$ we have $\Box f = (1/a^3(\eta))\, \Boxo (a(\eta) f)$. Hence
\be \int_{\Mo}\! \rmd^4 V \,\hphi(x)\, \Box f(x) = \int_{\Mo}\! \rmd^4 x \,\hphio(x)\,  \Boxo (af)\,(x)\, = \,0 \ee
where in the last step we have used the field equation (\ref{feo}) satisfied by $\hphio$ and the fact that $a\, f \equiv (a_1\,\eta) f(x) \in \S$. Finally, for the commutator we have 
\ba \label{relation} \hPhio (\Fo) &:=& \Omegao(\hphio(x),\, \Fo(x)) = \int_{\eta=\eta_{\circ}}\!\!\! \rmd^3 x \big[\hphio(x)\, \mathring{n}^a\nabla_a \Fo(x) - \Fo(x) \mathring{n}^a\nabla_a \hphio(x) \big] \nonumber\\
&=& \!\!\int_{\eta=\eta_{\circ}}\!\!\! \rmd^3 V \big[\hphi(x)\, n^a\nabla_a F(x) - F(x) n^a\nabla_a \hphi(x) \big] = \,\Omega(\hphi(x),\, F(x)) = \hPhi (F) \ea
where $\mathring{n}^a$ and $n^a$ are unit normals to any space-like plane $\eta=\eta_0 \not=0$ with respect to $\go_{ab}$ and $g_{ab}$ respectively,\, and $a(\eta) F(x) = \Fo(x)$. It then follows that in the extended FLRW space-time we have the commutation relations
\be [\hPhi(F),\, \hPhi (F^{\,\prime})]\, =\, i\hbar\, \Omega (F,\, F^{\,\prime})\, \h{I}   \ee
on $(\Mo, g_{ab})$. Since the symplectic inner product on the right can be defined as an integral over any surface $\eta=\eta_{\circ}\not=0$ and its value is independent of $\eta_{\circ}$, by continuity we can take the limit to the big bang surface $\eta=0$: although individual fields diverge, the volume element $\rmd^3 V$ shrinks to zero just in the right way to make the result well-defined (and non-zero). Using the relation $\hphi(x) = \hphio(x)/a(\eta)$ we can also evaluate the commutator between  OVDs directly:
\be [\hphi (x),\, \hphi(\xp)]\, =\, i\hbar\, \f{\Deltao(x,\, \xp)}{a(\eta) a(\eta^\prime)}\, \h{I} =: \, i\hbar\, \Delta(x,\, \xp)\, \h{I}. \ee
The commutator $\Delta(x,\, \xp)$ is a well-defined distribution on all of $(\Mo, g_{ab})$ and coincides with the distribution $(G_{\rm Ad} - G_{\rm Ret})$ if  $\eta, \eta^\prime$ are both positive or both negative. Its support is on the null cones of the extended manifold $\Mo$ (which, due to conformal flatness, are well-defined in spite of the singularity). Thus $\Delta(x,\, \xp)$ is a natural generalization of the distribution $(G_{\rm Ad} - G_{\rm Ret})$ across the singularity. To summarize, we have a well-defined  OVD $\hphi(x)$ on $(\Mo,g_{ab})$ satisfying all the requirements spelled out in section \ref{s2.1}. We will denote by $\A$ the $\star$-algebra generated by the smeared operators $\hphi(f)$ 
--or, equivalently by $\hPhi(F)$-- and by $\W$ the corresponding Weyl algebra.\medskip

Next, given any test function $f(x)$ on $\Mo$, we have
\be F(x):= \int_{\Mo} \! \rmd^4 V^\prime \Delta(x,\,\xp)\,f(\xp) = \f{1}{a(\eta)}\, \int_{\Mo} \! \rmd^4 \xp \Deltao(x,\,\xp)\, \fo(\xp)  :=   
\f{1}{a(\eta)}\, \Fo(x)\ee
where, as before $\fo(\xp) = a^3(\eta) f(\xp)$.  By its definition, $\Fo(x)$ is a smooth solution to the massless Klein-Gordon equation on $(\Mo, \go_{ab})$ and $F(x)$ is a distributional solution to the massless Klein-Gordon equation on $(\Mo, g_{ab})$. It follows from (\ref{relation}) that
\be \Omega(F_1(x),\, F_2(x))\, =\, \Omegao (\Fo_1(x),\, \Fo_2(x)).  \ee
Now, on Minkowski space $(\Mo, \go_{ab})$ we can expand out $\Fo(x) \in \pso$ as
\be \label{expansion1} \Fo(x) = \int \f{\rmd^3 k}{(2\pi)^3}\, \Big(z(\vk)\,\f{e^{-ik\eta}}{\sqrt{2k}}\, 
+ {z^\star}(-\vk)\,\f{e^{ik\eta}}{\sqrt{2k}}\Big)\, e^{i\vk\cdot\vx}\, , \ee
where the coefficients $z(\vk)$ belong to the Schwartz space $\t\S$; the complex structure $J$ on $\pso$ is given by
\be \Jo\,\Fo(x) = \int \f{\rmd^3 k}{(2\pi)^3}\, \Big(iz(\vk)\f{e^{-ik\eta}}{\sqrt{2k}}\, 
 -i{z^\star}(-\vk)\f{e^{ik\eta}}{\sqrt{2k}}\Big)\, e^{i\vk\cdot\vk}\, ; \ee
and the Hermitian inner product is 
\be \label{ip3} \langle \Fo_1,\, \Fo_2 \rangle_{\circ} = \f{1}{\hbar}\,\int \f{\rmd^3 k}{(2\pi)^3}\, z_1^\star(\vk)\, z_2(\vk)\, . \ee

It is natural to define $\ps$ (associated with $(\Mo, g_{ab})$) as the space of all $F(x)= \big(\Fo(x)/a(\eta)\big)$ with $\Fo(x) \in \pso$.  Then the complex structure on $\pso$ naturally induces a complex structure on $\ps$:\, $J F = (1/a(\eta))\, \Jo \Fo$. (Indeed, this is the complex structure normally used for massless scalar fields in the radiation filled FLRW space-time.) Thus, we have:
\ba F(x) &=& \int \f{\rmd^3 k}{(2\pi)^3}\, \Big(z(\vk)\f{e^{-ik\eta}}{\sqrt{2k}\,\,a_1\eta}\, + {z^\star}(-\vk)\f{e^{ik\eta}}{\sqrt{2k}\,\,a_1\eta}\, \Big)\, e^{i\vk\cdot\vx}\quad {\rm and,}\label{expansion2}\\
\langle F_1,\, F_2 \rangle &=& \f{1}{\hbar} \int \f{\rmd^3 k}{(2\pi)^3}\, z_1^\star(\vk)\, z_2(\vk)\, . \label{ip4}\ea
Consequently, although solutions $F(x) \in \ps$ diverge at $\eta=0$ as functions, each of them yields a  well-defined element of the 1-particle Hilbert space $\H$ on $(\Mo, g_{ab})$ because its norm is finite (and non-vanishing). This is analogous to the fact that, in quantum mechanics on $\mathbb{R}^3$, although the wave function $\psi(\vx) = {e^{-r}}/{r}$ diverges at $r=0$, it is nonetheless a well-defined quantum state, i.e., element of $L^2(\mathbb{R}^3)$. Finally, note  that if $f(x)\in \S$, then the corresponding solution $F(x)$ (of Eq.(\ref{sol1})) has finite norm with respect to the inner product (\ref{ip4}).

Given the Hermitian inner product (\ref{ip4}) on $\ps$, the Fock representation of the Weyl algebra $\W$ can be readily constructed using the GNS construction \cite{gns1,gns2} following the procedure summarized at the end of section \ref{s2.2}: the required positive linear function $\E$ on $\W$ is determined by the linear function $\e$ on $\ps$ as in (\ref{plf1})
\be \e(F) =  e^{{-\f{1}{2\hbar}\, \int \f{\rmd^3 k}{(2\pi)^3}\, {z^\star}(\vk)\, z(\vk)}}\, . \ee
As one would expect from (\ref{expansion1}) and (\ref{expansion2}), on this Fock space, the  OVD $\hphi(x)$ admits an explicit expansion in terms of annihilation and creation operators:
\be \hphi(x) = \f{1}{a_1\eta}\,\hphio(x) = \f{1}{a_1\eta}\, \int \f{\rmd^3 k}{(2\pi)^3}\, \Big(\Ak \f{e^{-ik\eta}}{\sqrt{2k}}\, + \Admk\f{e^{ik\eta}}{\sqrt{2k}}\, \Big)\, e^{i\vk\cdot\vx} .\ee
In spite of the $1/\eta$ term in the expression, as we saw in Eq.(\ref{ovd1}), $\hphi(f):= \int_{\Mo}\! \rmd^4 V \hphi(x) f(x)$ is a well-defined operator on the Fock space for all test fields $f$ in the Schwartz space $\S$.

\subsection{Expectation values and the Big Bang}
\label{s3.2}

Let us now examine the 2-point bi-distributions in the vicinity of the big bang, and then the renormalized expectation values of operator products that have received considerable attention in the literature because of their physical interest.

\subsubsection{The 2-point bi-distribution}
\label{s3.2.1}

Since $\hphi(f)$ is a well-defined operator on the Fock space $\F$ for all $f\in \S$, it is clear that the vacuum expectation value $\langle \hphi(f_1) \hphi(f_2)\rangle$ is also well-defined for all $f_1, f_2 \in \S$. The corresponding 2-point bi-distribution is given by
\ba \label{2point1} \langle \hphi(x)\, \hphi(\xp)\rangle &=& \f{1}{a_1^2 \eta\eta^\prime}\, \langle \hphio(x)\, \hphio(\xp)\rangle_{{}_{\circ}}\, =\, \f{\hbar}{a_1^2 \eta\eta^\prime}\, \int \f{\rmd^3 k}{(2\pi)^3}\, \f{e^{-ik t}}{2k}\,\, e^{i\vk\cdot\vr} \nonumber\\
&=& \f{\hbar}{a_1^2 \eta\eta^\prime}\, \f{1}{4\pi^2}\, \Big(\f{1}{r^2 - (t-i\epsilon)^2}\Big)\, , \ea
where we have set $t = \eta-\eta^\prime$ and $\vr = \vx -\vx{}^\prime$. The $i\epsilon$ prescription means that the integral --representing the Minkowski space 2-point bi-distribution-- should be regarded as the boundary value of an analytical function in the lower half of the complex $t$ plane. Thus even in Minkowski space, $\langle \hphio(x)\, \hphio(\xp)\rangle_\circ$ is not a function of 2 variables but a bi-distribution: one has to first integrate $\big(r^2 - (t-i\epsilon)^2\big)^{-1}$ against test functions and then take the limit $\epsilon \to 0$. Finally, although there is an overall multiplicative factor $1/\eta\eta^\prime$ in the expression of $\langle \hphi(x)\, \hphi(\xp)\rangle$, as we saw above, the final result is a well-defined bi-distribution on the entire $(\Mo, g_{ab})$ because of the $a^4(\eta)$ factor in the volume element $\rmd^4 V$. 

Although the Minkowskian $\langle \hphio(x)\, \hphio(\xp)\rangle_\circ$ is a bi-distribution, for points $x,\xp$ that are \emph{space-like or time-like} separated  one often considers the function $\f{\hbar}{4\pi^2} (\f{1}{r^2 - t^2})$ obtained by setting $\epsilon=0$ and interprets it as providing correlations between fields evaluated at points $x,\, \xp$ in the vacuum state. By inspection, this correlation function depends only on the space-time distance between $x$ and $\xp$ and falls off as the inverse power of the square of the distance, irrespective of whether the points are space-like separated or time-like separated. This is taken as a signature of the long range character of correlations of a massless scalar field in the Minkowski vacuum for both space-like and time-like separations. In the extended FLRW space-time $(\Mo, g_{ab})$, by contrast, there is an asymmetry between space and time. It is therefore of interest to probe how this asymmetry manifests itself in the correlation functions, especially near the big bang.

To carry out this task we have to make a choice of space-like and time-like separated points. We will use the simplest ansatz that is well-tailored to probe the asymmetry. We will take the time-like separated points to be $(\vx_{\circ}, \eta_{\circ})$ and $(\vx_{\circ}, \etap)$ such that $\eta^\prime >\eta_{\circ} > 0$ and the geodesic joining them --the straight line in the $(x_{\circ}, \eta)$ chart-- has proper length $D$. Thus $ (a_1/2)(\eta^\prime{}^2 - \eta_{\circ}^2) = D$. We will take the space-like separated points to lie on a $\eta=\eta_{\circ}$ surface such that the straight line connecting them --the geodesic within this surface-- has length $D$. Thus we will consider points $(\vx, \eta_{\circ})$ and $(\vxp, \eta_{\circ})$ such that $(a_1\eta_{\circ})\, |\vx - \vxp| \equiv (a_1\eta_{\circ})\, r = D$. The question is: How are the 2-point functions associated with these two pairs of points related? For space-like separated points we have:
\be \label{2point2} \langle \hphi(\vx, \eta_{\circ})\, \hphi(\vxp, \eta_{\circ} )\rangle = \f{\hbar}{4\pi^2 D^2} \ee
and, for the time-like separated points we obtain
\be \label{2point3} \langle \hphi(\vx_{\circ}, \eta_{\circ})\, \hphi(\vx_{\circ} \etap )\rangle = 
- \f{\hbar}{4\pi^2 a_1^2} \,\, \Big(\big(\f{2D}{a_1} + \eta_{\circ}^2\big)^{\f{1}{2}}\, \big(\eta_{\circ} - (\f{2D}{a_1} + \eta_{\circ}^2)^{\f{1}{2}}\big)^2 \Big)^{-1}\, \f{1}{\eta_{\circ}}\, .\ee
We will now take the limit as $\eta_{\circ} \to 0$, \emph{keeping the separation $D$ between the points fixed}. The first correlation function (\ref{2point2}) is $\eta_{\circ}$ independent while the second correlation function (\ref{2point3}) diverges as $1/\eta_{\circ}$. Therefore, for the ratio we have 
\be \label{ratio1} \lim_{\eta_{\circ} \to 0^+}\,\, \f{ \langle \hphi(\vx, \eta_{\circ})\, \hphi(\vxp, \eta_{\circ} )\rangle}{\langle \hphi(\vx_{\circ}, \eta_{\circ})\, \hphi(\vx_{\circ} \etap )\rangle\,}=\, -  \lim_{\eta_{\circ} \to 0^+}\,\,\f{2\sqrt{2 a_1}}{\sqrt{ D}}\, \eta_{\circ} = 0 \ee
where we have used the fact that $D$ is kept constant while taking the limit. Thus, 
in contrast to the situation in Minkowski space --where this ratio of correlation functions equals $1$ everywhere-- now the ratio vanishes as we approach the singularity, being proportional to $\eta_{\circ}$. For any given value of $\eta_{\circ}$, the coefficient $1/\sqrt{a_1 D}$ decreases as we increase the separations $D$; the asymmetry is enhanced as the separation between points increases since one probes a greater time dependent portion of space-time while calculating (\ref{2point3}). We can also consider the time-like separated points that lie on the two sides of the singularity by taking $\etap >0$ and $\eta=\eta_{\circ} <0$. The result (\ref{ratio1}) for the ratio is the same.

To summarize, the Minkowski space correlation function for two space-like separated points is the same as that for two time-like separated points if the physical distance between each pair is the same. In the radiation filled FLRW model, by contrast, the ratio (\ref{ratio1}) goes to zero as one approaches the singularity because, while the correlation function (\ref{2point2}) between space-like separated points is time independent if the distance between them is kept fixed, the correlation function (\ref{2point3}) between time-like separated points is time dependent --diverging as $1/\eta_{\circ}$ as $\eta_{\circ} \to 0$-- if, again, the distance between them is kept fixed. This asymmetry reminds one of the BKL behavior in classical general relativity \cite{bkl,berger}. However, we should keep in mind that the conceptual settings are very different. In the classical analysis, one investigates the behavior of the \emph{gravitational} field using the \emph{full non-linear Einstein equations}, and the BKL behavior refers to the fact that the time derivatives of the metric dominate over the spatial derivatives in an appropriate sense as one approaches space-like singularities. Here we considered \emph{test, quantum} fields on a cosmological background and found that correlations between time-like separated points dominate over those between space-like separated points as one approaches the singularity, even when the distance between points in each pair is kept fixed and equal.

\subsubsection{Renormalized operator products}
\label{s3.2.2}

So far, we have considered observables that can be constructed using the field operators $\hphi(f)$. The big bang singularity is tame if we probe it with these observables. We can also consider products of the type `$\hphi^2(x)$.' However, even in Minkowski space these observables are singular because $\hphi(x)$ itself is an OVD  \cite{asw}. But, as is well-known, one can renormalize their expectation values $\phisq$. One would expect these renormalized observables to provide sharper tools to probe the singularity because the renormalization procedure digs deeper into the ultraviolet behavior of the field. The most commonly used among them are the product $\phisq$ and the stress-energy tensor $\stress$. We will now investigate the behavior of both these observables near the big bang singularity in the extended FLRW space-time $(\Mo, g_{ab})$. 

As is well-known, the renormalization procedure involves subtracting a counter term that removes the `universal' part of the divergence. The standard strategy is to use a point splitting procedure \cite{bsd1,bsd2,smc1,smc2} and the DeWitt-Schwinger subtraction term \cite{bsd2}, constructed from curvature tensors of the background space-time metric.  There is extensive literature on calculations of $\phisq$ and $\stress$ that provides expressions of these observables as smooth fields in FLRW space-times, i.e. for $\eta >0$ or $\eta <0$ in our notation (see, in particular \cite{dfcb,bd1,ori,valencia} and references therein). Since they involve curvature tensors, they can diverge as $\eta\to 0$. The question we wish to address is whether these fields can nonetheless be regarded as well-defined distributions on $(\Mo, g_{ab})$.%
\footnote{\label{fn1} {Tempered distributions $\underline{\eta}^{-m}$ corresponding to the singular functions $\eta^{-m}$ are defined using successive derivatives of (the locally integrable function) $\ln |\eta|$ (just as the Dirac $\delta$ distribution and its derivatives are defined by derivatives of the step function even though it is not differentiable as a function). Appendix \ref{a1} discusses of these distributions and their properties.}}
(Recall that the classical solutions $F(x)$ of (\ref{expansion2}) diverge at $\eta=0$ but are nonetheless well-defined as distributions.)

Let us begin with $\phisq$. For definiteness, following \cite{ori,valencia} let us point-split along a space-translation Killing field. Then we have $t= \eta-\eta^\prime =0$  whence
\be \langle \hphi(x)\, \hphi(\xp) \rangle\, =\,  \frac{\hbar}{4\pi^2 a_1^2\eta\eta^\prime (r^2 -t^2)}\, \equiv\, \frac{\hbar}{4\pi^2 a_1^2\eta^2\, r^2} \ee
which of course diverges as we bring the points together, i.e., as $r \to 0$. Using the fact that $\h\phi$ is a massless scalar field and that the scalar curvature of the FLRW background vanishes, the expression of the DeWitt-Schwinger counter term \cite{dfcb} simplifies:
\be G_{\rm DS}(x,\xp)=\frac{\hbar}{8\pi^2\sigma} + \frac{\hbar}{96\pi^2\sigma}R_{ab}\sigma^a\sigma^b \ee
where $2\sigma$ is the signed square of the geodesic distance between $x$ and $\xp$ and $\sigma_a = \nabla_a \sigma (x,\xp)$. For our choice of $x$ and $\xp$, we have 
\be \sigma(x,\,\xp)=\f{1}{2}\, a^2(\eta)\, r^2\,+\, \mathcal{O}(r^4)  \quad {\rm whence} \quad  G_{\rm DS}(x,\xp)=\frac{\hbar}{4\pi^2a_1^2\eta^2 \,r^2} + \mathcal{O} (r^2)\,. \ee
Therefore, 
\be \phisq\,=\, \lim_{\xp\to x}\, \big[\langle\phi(x)\phi(\xp)\rangle\, - \, G_{\rm DS}(x,\, \xp) \big]\, =\, 0\, \ee
whence the answer to the question we set out to address is trivially in the affirmative. We can also choose to point-split in the time-like direction. Then the intermediate steps are a bit more complicated --in particular, the term $R_{ab} \sigma^a\sigma^b$ has a non-zero contribution in the $r\to 0$ limit-- but the final result is the same. Vanishing of $\phisq$ could be anticipated on dimensional grounds, since the only curvature scalar $S$ such that\, $\hbar S$  has physical dimensions of $\phisq$ is the Ricci scalar and it vanishes identically in the radiation filled FLRW space-time. (Recall that $\phisq$ is a dimension 2 observable.) We will find that the situation is non-trivial in the more general models with non-zero scalar curvature, but $\phisq$ continues to be a well-defined distribution on full $(\Mo, g_{ab})$.

Let us now consider $\stress$. It follows from the detailed calculations in \cite{dfcb} that for the massless scalar field in the radiation filled FLRW universe, 
we have 
\be \stress\, =\, \f{\hbar}{2880\pi^2}\, {}^{(3)}H_{ab} \quad {\rm where}\quad {}^{(3)}H_{ab}\, = \, R^{cd}R_{acbd}\,\ee
Using the explicit form $a(\eta)= a_1\eta$ we find 
\ba \label{SET1} \stress \,&=&\, \frac{\hbar} {720\pi^2 a_1^2\eta^6}\, \nabla_a\eta \nabla_b\eta + \frac{\hbar}{576\pi^2 a_1^2\eta^6}\, \go_{ab}\nonumber\\
\, &=& \, T_1(\eta)\, \nabla_a\eta \nabla_b\eta \,+\, T_2(\eta)\, \go_{ab}
\ea
where the 1-form $\nabla_a \eta$ and the Minkowski metric $\go_{ab}$ are well-defined on all of $\Mo$. (Note that we have the well-known trace anomaly.) Thus, even though the equation $\Box \hphi =0$ is conformally covariant, and $\phisq$ vanishes in the radiation filled FLRW space-time just as it does in Minkowski space-time, the observable $\stress$ senses the difference between the two space-times. The action of {$\stress$} on a generic test field ${f^{ab}(x)} = f_1(x) \eta^a \eta^b + f_2(x) \go^{ab}$ with $f_1(x),\, f_2(x) \in \S$ will be given by 
\be \stress:\, f^{ab}\, \rightarrow \, \int_{\Mo} \rmd^4 x\, (a_1^4\, \eta^4) (T_1f_1 + 4 T_2 f_2)\, .\ee   
Since $T_1$ and $T_2$ fall off as $1/\eta^6$, we are left with the action of the distribution  $1/\eta^2$ on the test functions $f_1$ and $f_2$. As explained in Appendix \ref{a1}, $1/\eta^2$ is a tempered distribution, which can be defined as the second derivative of the locally integrable function $\ln |\eta|$ in the distributional sense. 
To summarize, then, $\stress$ is a well-defined distributions on the extended space-time $(\Mo, g_{ab})$.\\

\emph{Remark:} Our main emphasis is on probing the big bang singularity with quantum fields from a mathematical physics perspective. Nonetheless, it is worth noting that 
the radiation filled FLRW space-time is also of physical interest since this model is sometimes taken to represent the universe before the onset of inflation (see, e.g., \cite{vm}).

\section{Quantum fields in a time dependent external potential}
\label{s4}
 
As we saw in Section \ref{s3}, certain technical simplifications arise in the radiation filled FLRW space-times because $a(\eta) = a_1 \eta$ implies that the scalar curvature $R$ vanishes. Therefore, we are naturally led to ask: Do the main features survive in more general FLRW models? As is well-known, in any FLRW model, the wave equation\, $\Box \phi =0$ \,can be simplified by setting $\chi(x) = a(\eta) \phi(x)$:\, Then $\chi(x)$ satisfies a wave equation in \emph{Minkowski space-time} but with a time dependent potential. Let us consider scale factors of the form $a(\eta) = a_\beta\, \eta^{\beta}$ with $\beta > 0$ so that we have a big bang/big crunch singularity at $\eta=0$. Then the FLRW metric is a solution to Einstein's equation with equation of state $p =w\rho$ with $w= \f{(2-\beta)}{3\beta}$ and $\chi(x)$ satisfies

\be \label{EOM3} \big(\Boxo + V(\beta, \eta) \big)\, \chi(x) =0, \quad {\rm where} \quad  V(\beta, \eta) = \f{\beta(\beta-1)}{\eta^2}\, \ee
whence the new basis functions\, $\eal(k, \eta):= a(\eta) e(k, \eta)$\, satisfy
\be \label{EOM4} \eal^{\prime\prime}(k, \eta) + \big(k^2 - \f{\beta(\beta -1)}{\eta^2}\big) \,\, {\eal}(k, \eta) =0. \ee
Note that the potential vanishes if $\beta=1$ --i.e. for the radiation filled universe we investigated in Section \ref{s3}-- thereby simplifying the analysis considerably. $\beta =0$ corresponds to Minkowski space. For values of $\beta$ other than these two, the $\eta$-dependence of the potential is universal; it diverges as $ \sim 1/\eta^2$ at $\eta=0$. Thus, we are naturally led to investigate the effect of this singularity in $V(\beta,\eta)$ on properties of the quantum field $\hchi (x)$ \emph{in Minkowski space-time} $(\Mo, \go_{ab})$. Now the volume element is just $\rmd^4 x$; \, it does not go to zero at $\eta =0$. Is $\hchi(x)$ nonetheless a well-defined operator-valued distribution on $(\Mo, \go_{ab})$? In this section, we will investigate this issue for $\beta =2$ and use these results in the next section to probe the big bang singularity in a dust filled universe using the quantum field $\hphi(x)$. For general values of $\beta$, the mode functions $\eal$ do not have a simple closed form, whence explicit calculations become quite involved, obscuring the underlying structure. However, as we discuss in Appendix \ref{a2}, the conceptual considerations in the more general case are very similar.

This section is divided in two parts. In the first we introduce quantum algebra $\Aoub$ generated by operators $\hchi(F)$ associated with suitably regular solutions $F$ to (\ref{EOM3}) and a natural Fock representation. In the second, we discuss the OVD $\hchi(x)$ and the 2-point bi-distribution $\bichi$. This construction encounters a mild infrared issue \cite{fp} and requires, as usual, an appropriate regularization. However, this issue is unrelated to the singularity in the potential at $\eta=0$; it exists even if we restrict ourselves to the manifold $M$ defined by $0<\eta <\infty$. Indeed, this issue arises also in de Sitter space-time (which corresponds to $\beta=-1$), where there is no singularity at all. 

\subsection{The quantum algebra $\Aoub$ and its quasi-free representation}
\label{s4.1}

Our discussion is divided into three parts. In the first we introduce the covariant phase space $\psoub$ of suitably regular solutions $\Foub(x)$ to (\ref{EOM3}); in the second we show that it admits a natural complex structure $\Jub$; and in the third we use the resulting K\"ahler space to construct a Fock representation of the quantum algebra $\Aoub$ generated by $\hTheta(F)$.

\subsubsection{The covariant phase space $\psoub$} 
\label{s4.1.1}
 
Since we have restricted ourselves to the $\beta=2$ case in this section, for notational simplicity we will denote the solutions $\eal(k, \eta)$ to (\ref{EOM4}) by $\eoub(k, \eta)$. Note that the equation satisfied by $\eoub(k, \eta)$ is exactly the same as that satisfied by $\eal(k, \eta)$ for $\beta=-1$ for which, as we noted above, the scale factor is the same as that in de Sitter space-time, $a(\eta) = 1/(H\eta)$ with $H^{-1} = a_{-1}$. Therefore, the well-known basis functions in de Sitter space
\be \label{basis1} \eoub (k, \eta) = \f{e^{-ik\eta}}{\sqrt{2k}}\, \big(1 - \f{i}{k\eta}\big) \ee 
and their complex conjugates provide us with two linearly independent solutions to (\ref{EOM4}) for each $k$. Each $\eoub( k, \eta)$ is singular as a function at $\eta=0$. However, it is a tempered  distribution on the full real line, defined using the Cauchy principal value of integrals (see Appendix A). Thus, 
$\eoub( k, \eta)$ is also a tempered distribution which, furthermore satisfies (\ref{EOM4}) in the distributional sense.%
\footnote{\label{fn2} More precisely, using properties (\ref{properties1}) of the tempered distributions $\underline{\eta}^{-m}$ it is straightforward to show that $\eoub(k,\eta)$ satisfies $\int_{\infty}^{\infty} \rmd \eta\,\, \eoub(k,\eta) (\mathcal{O} f)=0$ for all $f$ such that $\mathcal{O} f$ is in $\S$. Here, $\mathcal{O} = (\rmd^2/\rmd \eta^2 + k^2 - (2/\eta^2))$ is the symmetry reduced `wave operator' (see (\ref{EOM4})).}
Therefore, we can follow the procedure of Section \ref{s2.3} to construct the covariant phase space $\ps$. It will consist of solutions $\Foub(x)$ to (\ref{EOM3}) of the type
\be \label{expansion3} \Foub(x) = \int \f{\rmd^3 k}{(2\pi)^3}\, \Big(z(\vk)\, \eoub (k,\eta) + {z^\star}(-\vk)\, \eoub^\star(k,\eta)\, \Big)\, e^{i\vk\cdot\vx} \ee 
where $z(\vk)$ belong to the Schwartz space $\t\S$ associated with the 3-dimensional momentum space. Note that although the expression of $\eoub (k, \eta)$ contains a term that diverges as $1/k^{3/2}$ at $k=0$, the integral is infrared finite because of the $k^2$ factor in $\rmd^3 k$. The expression also contains a term that diverges as $1/\eta$ at $\eta=0$. However, for reasons explained above, each $F(x)$ is a well-defined tempered distribution on Minkowski space $(\Mo, \go_{ab})$ that satisfies the field equation (\ref{EOM3}) in the distributional sense.

Next, let us evaluate the symplectic inner product between two elements $\Foub_1(x)$ and $\Foub_2(x)$ in $\psoub$. Since the symplectic current is conserved, we can evaluate the symplectic structure using any Cauchy surface, say $\eta=\eta_\circ \not=0$. Using the unit normal $\no^a$ to the Cauchy surface, we have
\be \label{symp1} \Omegaoub\,(\Foub_1, \, \Foub_2) = \int_{\eta=\etao}\!\! \rmd^3 x\, \big(\Foub_1\, \no^a\nabla_a \Foub_2\, -\, \Foub_2\, \no^a\nabla_a \Foub_1\big)\, .\ee 
Let us substitute the expression (\ref{expansion3}) of $\Foub_1(x)$ and $\Foub_2(x)$ in (\ref{symp1}) and simplify. Since $\eoub (k, \eta_\circ)$ has a term that goes as $1/k^{3/2}$ the integrand of the symplectic structure has terms that go as $1/k^3$ that would be infrared divergent. However, because of anti-symmetry between $\Foub_1(x)$ and $\Foub_2(x)$ these terms cancel. More precisely, in the intermediate step, we will replace the integral\, $\int \rmd^3 k (...)$\, by\,\, $\h{\!\int} \rmd^3 k (...) := \lim_{\ell \to \infty}\, \int_{1/\ell}^\infty \rmd k\,k^2 \oint \rmd^2 \Omega_{\vk}\, (...)$, \, carry out the cancellation in the truncated integral and then take the limit. We obtain:
\ba \label{symp2}\Omegaoub\,(\Foub_1, \, \Foub_2) &=& \widehat{\hskip-.25cm\int} \f{\rmd^3k}{2\pi^3}\, \big[ \big(z_1(\vk) \eoub (k,\eta_\circ) + z_1^\star(-\vk) \eoub^\star (k,\eta_\circ)\big) \big(z_2(-\vk) \eoub^\prime (k,\eta_\circ) + z_2^\star(\vk) \eoub^{\prime\,\star} (k,\eta_\circ)\big)\,  -\, 1 \leftrightarrow 2\, \big ]\nonumber\\
&=& \int \f{\rmd^3k}{2\pi^3}\, \big[z_1(\vk) z_2^\star (\vk) - z_2(\vk) \bar{z}_1^\star (\vk)\big]\, \big(\eoub (k, \eta_\circ) \eoub^{\prime\, \star} (k, \eta_\circ) - \eoub^\star (k, \eta_\circ) \eoub^{\prime} (k, \eta_\circ)\big) \nonumber\\
&=& i\,\int \f{\rmd^3k}{2\pi^3}\, \big[z_1(\vk) z_2^\star(\vk)\, - \, z_2(\vk) z_1^\star(\vk)\big]\ea
where, in the last step we have used the normalization condition
\be \eoub (k, \eta_\circ) \,\eoub^{\prime\, \star} (k, \eta_\circ)\, -\, \eoub^\star (k, \eta_\circ)\, \eoub^{\prime} (k, \eta_\circ)\, =\, i \ee
satisfied by the functions $\eoub(k,\eta)$. Since $z_1(\vk)$ and $z_2(\vk)$ are in $\t\S$, the final integral in (\ref{symp2}) is well-defined. Thus, as in Section \ref{s3.1}, even though the solutions $F(x)$ are distributional, $\psoub$ carries a well-defined symplectic structure. (Note, incidentally, that the infrared regularization that is necessary in an intermediate step has nothing to do with the distributional character of $F(x)$ since the regularization is needed on the $\eta=\eta_\circ \not=0$ surface in a neighborhood of which $F(x)$ is smooth.) 
 
\subsubsection{A natural complex structure\,\,\, $\mathring{\!\!\!\underbar{J}}$ on $\psoub$}
\label{s4.1.2} 
  
We will now show that $\psoub$ carries a natural complex structure\, $\Joub$,\, thanks to the fact that the potential  $V(\eta)$ decays sufficiently fast as $\eta \to \pm\infty$ (and is symmetric under $\eta \to - \eta$). 
 
We begin with two observations. First, given a time instant $\etao$, one can set a natural isomorphism $I_{\etao}$ from the phase space $\psoub$ of solutions $\Foub(x)$ to (\ref{EOM3}) and the phase space $\pso$ of solutions $\Fo(x)$ to the massless Klein-Gordon equation in Minkowski space using initial data:\, $I_{\etao} \Foub(x) = \Fo(x)$ if and only if\,  $\Foub(\vx, \etao) = \Fo(\vx, \etao)$\, and\, $\Foub^\prime(\vx, \etao) = \Fo^\prime(\vx, \etao)$,\, where as usual the prime denotes derivative with respect to the Minkowskian $\eta$. Second, given \emph{any} space-like plane $\eta= \etao$, the standard Minkowskian complex structure $\Jo$ on $\pso$ has the following action on initial data:
\be \Jo (\Fo(\vx, \etao), \,\Fo^\prime(\vx, \etao))\, =\, (-{\mathring\Delta}^{-\f{1}{2}} \Fo^\prime (\vx, \etao),\,\, \mathring\Delta^{\f{1}{2}} \Fo (\vx,\etao))\, ,\ee 
where $\mathring\Delta$ is the 3-dimensional, flat Laplacian. Therefore, there is a natural a 1-parameter family of complex structures, $\Jub_{\etao} = I_{\etao}^{-1}\, \Jo\, I_{\etao}$ on $\psoub$. Setting
\be \Jub_{\etao} \Foub(x)\, =\,  \t{\Foub}(x)\, = \int \f{\rmd^3 k}{(2\pi)^3}\, \Big(\t{z}(\vk)\, \eoub (k,\eta) + {\t{z}^\star}(-\vk)\, \eoub^\star(k,\eta)\, \Big)\, e^{i\vk\cdot\vx}    
\ee
one finds 
\be \label{Jetao} \t{z}(\vk) = i \Big[ \f{|\eoub^{\prime\, \star}(k,\etao)|^2}{k} + k |\eoub(k, \etao)|^2\Big]\, z (\vk) \,+\,  i \, \Big[\f{(\eoub^{\prime\, \star}(k,\etao))^2}{k} + k (\eoub(k, \etao))^2\Big]\, z^\star (-\vk)\ee
Thus, while on the Minkowskian phase space $\pso$, the complex structure $\Jo$ just multiplies $z(\vk)$ by $i$, on $\psoub$, the complex structure $\Jub_{\etao}$ sends $z(\vk)$ to an $\etao$-dependent complex-linear combination of $z(\vk)$ and $z^\star(\vk)$. Using the explicit expression (\ref{basis1}) of $\eoub(k)$, we obtain 
\be \t{z}(\vk) = i \Big[1 +\f{1}{2k^4\etao^4}\Big]\, z(\vk)\, +\, i \Big[\f{e^{2ik\etao}}{k^2\etao^2}\Big]\, \Big[1 + \f{i}{k\etao} + \f{1}{2k^2\etao^2}\Big]\, z^\star(-\vk)\, .\ee
Therefore, in the limit $\etao \to \pm \infty$\, --in which the potential $V(\eta) = -2/\eta^2$ goes to zero-- \,the complex structures $\Jub_{\etao}$ admit a simple limit\, $\Joub$ on $\psoub$:
\be \label{J2} \Joub\, \Foub(x)= \int \f{\rmd^3 k}{(2\pi)^3}\, \Big(i\,z(\vk)\, \eoub (k,\eta) \,-i\, {z^\star}(-\vk)\, \eoub^\star(k,\eta)\, \Big)\, e^{i\vk\cdot\vx}\, . \ee 
This construction has three noteworthy features: (i) The limits $\etao\to \pm \infty$ of $\Jub_{\etao}$ are well defined; (ii) The two limits provide the \emph{same} complex structure on $\psoub$; and, (iii) This complex structure $\Joub$ is compatible with the symplectic structure on $\psoub$. Thus, $\psoub$ admits a natural K\"ahler structure. The resulting Hermitian inner product on $\psoub$ is given by 
\be \label{ip5} \langle \Foub_1,\, \Foub_2 \rangle\, =\, \f{1}{\hbar} \int \f{\rmd^3 k}{(2\pi)^3}\, z^{\star}_1(\vk)\, z_2(\vk)\, , \ee 
which is formally the same expression as in the radiation filled universe. We could have just posited the complex structure $\Joub$ of (\ref{J2}) and the inner product (\ref{ip5}) on $\psoub$ as a natural generalization of the result in Section \ref{s3}. The construction we used brings out the fact that $\Joub$ arises naturally by considering the $\eta \to \pm\infty$ limits, in spite of the presence of a time dependent potential. Although one could have heuristically anticipated the result from the form of the basis functions $\eoub(k,\eta) = \big(e^{-ik\eta}/\sqrt{2k}\big) \big(1 - i/k\eta\big)$ since the second factor goes to $1$ in the limit $\eta \to \pm\infty$, one cannot draw a definitive conclusion from this fact since the phase factor in the first factor also oscillates uncontrollably in these limits.\\

\emph{Remark:} Note that Eq. (\ref{Jetao}) is not tied to the choice $\beta =2$ in (\ref{EOM4}); it holds for any $\beta$. Also, the argument that led us to the final form of the complex structure (\ref{J2}) depends only on the asymptotic forms of the basis functions $\eoub(k, \eta)$. These facts will be used in Appendix \ref{a2} in our discussion of more general FLRW space-times. 

\subsubsection{The Fock representation determined by\,\,\, $\mathring{\!\!\!\Jub}$.} 
 \label{s4.1.3}
 
Starting with the phase space $\psoub$, as before we can introduce the algebra $\Aoub$ generated by abstractly defined operators $\hTheta (F)$ that are linear in $\Foub$ and satisfy the commutation relations
\be [\hTheta (\Foub_1),\, \hTheta(\Foub_2) ] \, =\, i\hbar\,\, \Omegaoub \,(\Foub_1,\, \Foub_2)\, \h{I}\, . \ee
The complex structure $\Joub$ then leads us to the Fock representation of $\Aoub$ as before. The 1-particle Hilbert space $\Houb$ is the Cauchy completion of $\psoub$ under the inner product (\ref{ip5}). Thus, for $\Foub\in \Houb$, the $z(\vk)$ have to be only square integrable; they need not be smooth nor fall off faster than any polynomial. The carrier space of the representation is the symmetric Fock space $\Fockoub$ generated by $\Houb$ and the abstract operators $\hTheta(\Foub)$ are again represented as linear combinations of concrete, annihilation and creation operators on $\Fockoub$:
\be \hTheta(\Foub) = \hbar \big(\h{A}(\Foub) + \h{A}^\dag (\Foub) \big)\, .  \ee
However, as we will now see, a subtle infrared issue arises in defining the operator-valued distribution $\hchi(x)$ on $\Fockoub$, which was absent in our discussion of the radiation filled Friedmann universe in Section \ref{s3}.
 
\subsection{The operator-valued distribution $\hchi(x)$} 
\label{s4.2}
 
This section is divided into two parts. In the first we will discuss the infrared issue and in the second we will investigate the 2-point bi-distribution $\bichi$. The infrared issue arises already in the $\eta >0$ or $\eta <0$ parts of Minkowski space $(\Mo,\go_{ab})$, where the potential $V(\eta)$ is regular. Once it is handled,  $\hchi(x)$ is a well-defined operator-valued distribution on all of $(\Mo,\go_{ab})$; the fact that the potential is singular at $\eta=0$ does not create any new obstructions. 

\subsubsection{The infrared issue}
\label{s4.2.1}

Motivated by considerations of Section \ref{s2}, let us consider a putative operator-valued distribution
\be \label{putative} \hchi(x) = \int \f{\rmd^3 k}{(2\pi)^3}\, \Big(\h{A}(\vk)\, \eoub (k,\eta) + {\h{A}^\dag}(-\vk)\, \eoub^\star(k,\eta)\, \Big)\, e^{i\vk\cdot\vx}\, . \ee 
The question we wish to address is whether operators $\hchi(f)$ smeared with test functions $f$ are well-defined on the Fock space $\F$ of Section \ref{s4.1.3} for all $f\in \S$. Since annihilation and creation operators in $\F$ are associated only with solutions that have finite norm in the 1-particle Hilbert space $\H$ the question reduces to: Does the solution $\Foub$ determined by every test function $f(x)\in \S$ have finite norm in $\H$? To address this question let us first recall that the operators $\hchi (f)$ and $\hTheta(\Foub)$ are related by:
\be  \hchi(f) = \Omega \big(\hchi(x), \, \Foub(x)\big) =:  \h\Theta(\Foub) \ee
To make the structure of the argument transparent, let us first consider test functions of the type $f(x) = g(\eta) h(\vx)$ although it will be clear from the discussion that this restriction does not play an essential role. Then,
\ba \hchi(f) &=& \int \f{\rmd^3 k}{(2\pi)^3}\,\big[ \h{A}(\vk) \int\! \rmd^3x\, h(\vx) e^{i\vk\cdot\vx}\, \int\! \rmd\eta\,\, g(\eta) \eoub(k, \eta)\, +\,\, {\rm HC}\big]\nonumber\\
&=& \int \f{\rmd^3 k}{(2\pi)^3}\,\big[\h{A}(\vk) \tilde{h}(\vk)\, \int\!\! \rmd\eta \,\,\, g(\eta)\,\f{e^{-ik\eta}}{\sqrt{2k}}\, \big(1 - \frac{i}{k\eta}\big)\, +\,\, {\rm HC}\, \big]\, ,\ea
where $\t{h}(\vk)$ is the Fourier transform of $h(\vx)$ and hence in the Schwartz space $\t\S$ associated with the momentum space. Let us consider the two terms under the $\eta$-integral and set:
\be \label{etaintegrals}\t{g}^\star(k) := \int_{-\infty}^{\infty}\!\!\rmd \eta\, g(\eta)\, e^{-ik\eta} \qquad {\rm and} \qquad \t{I}^\star_g(k) :=  \int_{-\infty}^{\infty}\!\!\rmd \eta\,\, \f{1}{\eta}\, \big(g(\eta)\, e^{-ik\eta}\,\big) \ee 
where $\t{g}(k)$ is the Fourier transform of $g(\eta)$ in the 1-dimensional Schwartz space and $\t{I}_g(k)$ is the result of the action of the 1-dimensional tempered distribution $1/\eta$ on a test function $g(\eta)\,e^{ik\eta}$. (For a discussion of tempered distributions corresponding to $\eta^{-m}$ for any integer $m$, see Appendix \ref{a1}.) Then the expression of $\hchi(f)$ takes the form:
\be \hchi(f)= \int \f{\rmd^3 k}{(2\pi)^3}\,\Big[\, \t{h}(\vk)\, \big(\f{\t{g}^\star(k)}{\sqrt{2k}}\, -\, \f{i}{\sqrt{k^3}}\, \t{I}^\star_g(k)\big)\, \h{A}(\vk) \,+\, {\rm HC} \,\Big]
\ee
%
%
On the other hand, we also have:
\be \h\Theta(\Foub) = \int \f{\rmd^3 k}{(2\pi)^3}\,\Big[ \big(iz^\star(\vk)\big) \h{A}(\vk)\, + \, {\rm HC}\,\, \Big]\, . \ee 
Therefore, the putative solution $\Foub(x)$ determined by the test function $f(x)$ is characterized by
\be \label{z} z(\vk)\, = \, i \t{h}^\star(\vk)\, \Big(\f{\t{g}(k)}{\sqrt{2k}}\, + \, \f{i}{\sqrt{k^3}}\, \t{I}_g(k)\Big)\, . \ee
Recall from (\ref{ip5}) that $\Foub(x)$ belongs to the 1-particle Hilbert space $\H$ if and only if $z(\vk)$ is square integrable. Let us first check the ultraviolet behavior. Since $\t{h}^\star (\vk) \t{g}(k)$ falls off faster than any polynomial, the first term is clearly tame in the ultraviolet. Next, note that the definition of $\t{I}^\star_g$ implies that it is the anti-derivative with respect to $k$ of the function $-i\, \t{g}^\star(k)$ in the Schwartz space. (The notion of anti-derivative of distributions is recalled in Appendix \ref{a1}). Therefore we have 
\be \label{Istar} \t{I}^\star_g (k) = -i\, \int_{0}^{k}\!\! \rmd k^\prime\, \t{g}^\star(k^\prime)
\, +\, C_g \ee
where the constant $C_{g}$ is determined by the action of the distribution $1/\eta$. Therefore $\t{I}_g(k)$ is a smooth and bounded function of $k$, whence the second term is also ultraviolet tame. However,  because of the $1/\sqrt{k^3}$ term in front of $\t{I}^\star_g(k)$ in (\ref{z}), there is a potential infrared problem with square-integrability of $z(\vk)$. Now, if $C_g$ were to vanish, since $\t{I}^\star_g (k)$ is smooth, it would vanish (at least) as $\sim k$ at $k=0$ and then $z(\vk)$ would be square-integrable. Thus, $\hchi(f)$ is well-defined for $C_g =0$. Translating in terms of the original test function $f(x)$, we have: the operator $\hchi(f)$ is well-defined (i.e. without an infrared divergence) if and only if the spatial Fourier transform $\t{f}(\vk, \eta)$ of $f(x)$ satisfies: $\int \rmd\eta\, (1/\eta) \t{f}(\vec{0}, \eta) =0$. This is a single (linear) condition on test functions. Let us denote by $\S_1$ the subspace of $\S$ spanned by the $f(x)$ satisfying this condition. $\S_1$ is of co-dimension $1$ in $\S$ and the action of $\hchi(f)$ is well defined on the Fock space $\Fockoub$ for every $f\in \S_1$. In the language of distribution theory, we need to provide a single input to extend this action of $\hphi(x)$ to full $\S$: specify (or, regulate) the $z(\vk)$ defined by \emph{any one} test functions $f(x)$ that fails to lie in $\S_1$ (see Appendix \ref{a1}). In practice, this is done by introducing an infrared cut-off $1/\ell$; but the cut-off can be removed simply by taking the limit $\ell\to \infty$ for all $f(x) \in \S_1$.

To summarize, as we saw in Section \ref{s4.1}, we have a Fock representation of the algebra $\Aoub$ generated by operators $\hTheta (\Foub)$. The smeared field operators $\hchi(f)$ have a well-defined action on this representation if $f\in \S_1$, a (infinite dimensional) subspace of $\S$ with co-dimension $1$. To extend this action for all $f\in \S$, we only need to provide an additional input which however, is captured in a single parameter because the co-dimension of $\S_1$ in $\S$ is $1$. Note that this additional input is required because of a (well-controlled) infrared divergence that arises already for $\eta >0$ --the issue is unrelated to issues stemming from the blow-up of the potential $V(\eta)$ at $\eta=0$.

\subsubsection{The 2-point bi-distribution}
\label{s4.2.2} 
 
Let us now consider the putative bi-distribution $\bichi$ constructed from (\ref{putative}): 
\be \label{bichi1} \bichi = \hbar\, \int \!\f{\rmd^3 k}{(2\pi)^3}\,\, \f{1}{2k}\,\, e^{-ik(\eta-\eta^\prime)\, +\, i \vk\cdot(\vx -\vxp)}\,\, \big(1 - \f{i}{k\eta}\big)\big(1+ \f{i}{k\etap} \big)  \ee
By performing the angular integral, we obtain
\ba \label{bichi2} \bichi &=& \f{\hbar}{4\pi^2 r}\, \int_{0}^\infty \rmd k \, e^{-ikt}\, \sin kr\, \big(1 +\f{it}{k\eta\etap} + \f{1}{k^2\eta\etap}\big) \nonumber\\
&=& \f{\hbar}{4\pi^2 r}\,\,  \big({\rm I} + \f{it}{\eta\etap}\,\, {\rm II} + \f{1}{\eta\etap}\,\, {\rm III}\big)\ea
where, as before, $t= \eta-\etap$ and $r = |\vx-\vxp|$. The integrals I, II and III  are ill-defined in the ultraviolet and one has to regulate them by the $i\epsilon$ prescription $t \to t-i\epsilon$. (As we saw in Section \ref{s3.2} this is already needed for the field $\hphio$ in Minkowski space without any potential.) Then, we obtain three distributions. First, as in (\ref{2point1}), we have:
\be {\rm I}\, =\, \f{r}{r^2 - (t-i\epsilon)^2}\,\, \equiv  \,\, \Big(\f{1}{r-(t-i\epsilon)}\, +\, \f{1}{r+(t-i\epsilon)}\Big)\, . \ee
Second, $-i{\rm I}$ is the derivative of ${\rm II}$ with respect to $t$ whence, to obtain ${\rm II}$ we have to calculate the anti-derivative of the distribution ${\rm I}$ (see Appendix \ref{a1}). Integrating $-i \,{\rm I}$ with respect to $t$ (using the last expression of ${\rm I}$) we obtain: 
\be {\rm II}\, = \, \f{i}{2}\,\, \ln \f{|r - (t-i\epsilon)|}{|r +  (t-i\epsilon)|} \, \equiv\,   \f{i}{2}\,\,\Big[ \ln \f{|r - (t-i\epsilon)|}{\ell}\, -\, \ln \f{|r + (t-i\epsilon)|}{\ell}\Big]\ee
where in the last step we have introduced $\ell$ for dimensional reasons; the right side is independent of the choice of $\ell$. Note that the integral $\int \rmd k\, e^{-ikt}\, (\sin kr /k)$ defining ${\rm II}$ is infrared finite because of the $\sin kr$ term in the numerator (and all integrals are ultraviolet finite because of the $i\epsilon$ term). The infrared issue we encountered in Section \ref{s4.2.1} arises in the evaluation of the third term because now the integrand contains $\sin kr/k^2$ \cite{fp}. However, since $-i{\rm II}$ is the derivative of ${\rm III}$ we can evaluate ${\rm III}$ by integrating the last  expression of $-i {\rm II}$ with respect to $t$. In this procedure $\ell$ now serves as the infrared regulator. Adding the three terms, one obtains the bi-distribution:
\be \label{bichi3} \bichi = \f{\hbar}{4\pi^2} \,\,\Big[ \f{1}{r^2 - (t-i\epsilon)^2} \, - \, \f{1}{2\eta\etap} \ln \f{r^2 - (t-i\epsilon)^2}{\ell^2} + \f{1-\gamma}{\eta\etap}\Big] \ee
where, we have followed the conventions in the literature to fix the integration constants at each of the two steps (whence the presence of the Euler-Mascheroni constant $\gamma$ in the last term; it can be absorbed in $1/\ell^2$). As before, the $i\epsilon$ prescription means that we have to first carry out the integral against test functions and then take the limit as $\epsilon \to 0$. Because of the presence of $1/\eta\etap$ terms (as well as the very first term which is present already in Minkowski space without a potential!), the right side is singular as a function of $x$ and $\xp$. However, $\bichi$ is well-defined as a bi-distribution (since that $1/\eta$ is a tempered distribution on the full $\eta$-real line). \medskip

\emph{Remark:} If we specialize to $t=0$ --i.e., let $x$ and $\xp$ lie on a $\eta = \eta_{\circ} \not=0$ surface-- one can use the well-known fact that  in 3-dimensions the Fourier transform of $1/k$ is $1/(2\pi^2\, r^2)$ in the distributional sense to carry out the calculation using the 3-dimensional integral in (\ref{bichi1}) directly, without having to perform the angular integral that led us to (\ref{bichi2}).  This short calculation provides an independent check on (\ref{bichi3}), albeit for a special case.

\section{The dust filled universe}
\label{s5}

In this section we return to quantum fields in FLRW space-times. In the radiation filled universe considered in Section \ref{s3}, one might be tempted to say that the tameness of quantum fields $\hphi(f)$ and $\hPhi(F)$ simply trickles down from the regularity of $\hphio(x)$ in Minkowski space-time because of the conformal covariance of \,$\Box \hphi (x) =0$ \, in this FLRW space-time (although, as we pointed out, it is not a priori obvious that $\hphi(x) = a^{-1}(\eta\,)\hphio(x)$ should be tame in spite of the fact that $a(\eta)$ vanishes at the big bang). Therefore, it is useful to investigate whether this tameness continues to hold in more general FLRW space-times. In the dust filled universe --where $a(\eta) = a_{2}\eta^2$-- the scalar curvature does \emph{not} vanish, whence one cannot use the conformal covariance argument to relate $\hphi$ to the Klein Gordon field $\hphio$ satisfying $\Boxo \hphio =0$ in Minkowski space. Do the quantum fields $\hphi(f)$ and $\hPhi(F)$ still continue to be tame across the big bang (and big crunch) singularity? We will now show that the answer is in the affirmative. We will find that the tameness is in fact somewhat enhanced for the renormalized products of operators. 

This section is divided into three parts. In the first we investigate the field operators; in the second, the bi-distribution $\biphi$; and in the third, the renormalized products $\phisq$ and $\stress$.

\subsection{Field operators}
\label{s5.1}

Although solutions $F(x)$ to the wave equation in the dust filled FLRW space-time are not simply related to the solutions $\Fo(x)$ of the wave equation in Minkowski space-time, as we pointed out in the beginning of Section \ref{s4}, they \emph{are} simply 
related to the solutions $\Foub(x)$ to the wave equation (\ref{EOM3}) in Minkowski space-time, now in presence of a time dependent potential $V(\eta) = -2/\eta^2$:\, $F(x):= \Foub(x)/a(\eta)$ satisfies $\Box F =0$ (in a distributional sense) if and only if $\Foub(x)$ satisfies $(\Box - V(\eta))\,\Foub(x) =0$ (in a distributional sense). Therefore, it is easy to analyze properties of $\hPhi(F)$ and $\hphi(f)$ using results on $\hTheta(\Foub)$ and $\hchi(f)$ from Section \ref{s4}. The arguments are completely parallel to those we used in Section \ref{s3} to draw conclusions about $\hPhi(F)$ and $\hphi(f)$ from properties of $\hPhi(\Fo)$ and $\hphio(f)$ in Minkowski space-time. Therefore, our discussion will be rather brief.

In view of our discussion in Section \ref{s4.1.2}, is natural to use $e (k, \eta) := \eoub (k, \eta)/ a(\eta)$ as the (`positive-frequency') basis functions. Then covariant phase space $\ps$ will consist of solutions $F(x)$ to the wave equation in the FLRW space-time of the form
\be \label{expansion4} F(x) = \int \f{\rmd^3 k}{(2\pi)^3}\, \Big(z(\vk)\,  e(k,\eta) + {z^\star}(-\vk)\, e^\star(k,\eta)\, \Big)\, e^{i\vk\cdot\vx} \ee 
where, as before, $z(\vk)$ belong to the Schwartz space $\t\S$ associated with the 3-dimensional momentum space. It is easy to verify that the symplectic structure is again given by (\ref{symp2}) and the Hermitian inner product by (\ref{ip5}) (with the obvious replacement of $\Foub_1$ and $\Foub_2$ by $F_1$ and $F_2$). Thus,\, $F(x) \to \Foub(x) = a(\eta) F(x)$ is a natural isomorphism from the phase space\, $\ps$\, to the phase space\, $\psoub$\, of Section \ref{s4}. In particular, the symplectic structure $\Omega$ is the pull-back of the symplectic structure  $\Omegaoub$:\,
$\Omega(F_1,\, F_2) = \Omegaoub\, (\Foub_1, \, \Foub_2)$. Therefore, the Fock space $\Fockoub$ of Section \ref{s4} carries a natural representation of the algebra $\A$ generated by the field operators $\Phi(F)$:
\be \hPhi (F) = \h\Theta(\Foub) \qquad \hbox{so that}\qquad  [\hPhi (F_1),\, \hPhi (F_2)] = i\hbar\, \Omega (F_1,\,F_2)\, \h{I} \ee
as in Section \ref{s3.1}. Finally, the operator-valued distribution $\hphi(x)$ is represented on $\Fockoub$ as
\be \hphi(x) = \f{1}{a(\eta)}\,\hchi (x)\,=\, \int \f{\rmd^3 k}{(2\pi)^3}\, \Big(\h{A}(\vk)\,  e(k,\eta) + {\h{A}^\dag}(-\vk)\, e^\star(k,\eta)\, \Big)\, e^{i\vk\cdot\vx}\,  \ee 
again as in section \ref{s3.1}. However, now the infrared subtlety we found in Section \ref{s4} trickles down to $\hphi(x)$ as follows. We have
\be \label{ovd2} \int_{\Mo}\! \rmd^4 V\, \hphi(x) f(x)\, =\,  \int_{\Mo}\! \rmd^4 x \,\,\hchi(x)\, (a^3(\eta) f(x))\, \equiv \int_{\Mo}\! \rmd^4 x\,\, \hchi(x) f_\circ(x)\, , \ee
and $f_\circ(x) = a^3(\eta)f(x) $ is guaranteed to be in $\S$ if $f(x)$ is. Now, if $f_\circ(x) \in \S_1$\, --i.e., if $\int \rmd\eta\, (1/\eta)\, \t{f}_\circ(\vec{0}, \eta) =0$, where $\t{f}_\circ(\vk, \eta)$ is the Fourier transform  in $\vx$ of $f(\vx, \eta)$--\, the operator $\hphi(f)$ is well-defined on $\Fockoub$ as is. However, if it is not, we need an infrared regulator $\ell$. The key point is that the presence of the big bang (or big crunch) singularity  does not cause any additional complication because the volume element (more than) compensates the $1/\eta$ factor in $e(k,\eta)$.

\subsection{The bi-distribution $\biphi$} 
\label{s5.2}

With the infrared regulator in place $\hphi(x)$ is a well-defined OVD on $\Foub$. It then trivially follows that  
\ba \label{2point4} \biphi &=& \f{1}{a(\eta) a(\etap)}\, \bichi\, \nonumber\\
&=& \f{\hbar}{4\pi^2}\, \f{1}{a_{2}^2\, \eta^2 \eta^{\prime\,2}}\,\,\Big[ \f{1}{r^2 - (t-i\epsilon)^2} \, - \, \f{1}{2\eta\etap} \ln \f{r^2 - (t-i\epsilon)^2}{\ell^2} + \f{1-\gamma}{\eta\etap}\Big]\, . \ea
is a bi-distribution on $(\Mo, g_{ab})$. As in Section \ref{s3}, it is well-defined in spite of the presence of inverse powers of $\eta$ and $\eta^{\prime}$ because of the factor $a^4(\eta) = a_{2}^4\, \eta^8$ in the volume element of $g_{ab}$ relative to that of $\go_{ab}$.

As discussed in Section \ref{s3}, for points $x$ and $\xp$ that are space-like or time-like separated, the bi-distribution is often used as a measure of correlations between fields evaluated at two points. Therefore, it is of interest to investigate the asymmetry between these correlations between space-like and time-like separations in the vicinity of the big-bang singularity. The difference between the radiation filled and dust filled universe lies in the last two terms in the square bracket (and of course the fact that $a(\eta)$ now goes as $\eta^2$ rather than as $\eta$). We will find that these extra terms make a qualitative difference near the singularity.

We will again use the simplest ansatz that is well-tailored to probe the asymmetry. We will take the time-like separated points to be $(\vx_{\circ}, \eta_{\circ})$ and $(\vx_{\circ}, \etap)$ such that $\eta^\prime >\eta_{\circ} > 0$ and the geodesic joining them --the straight line in the $(x_{\circ}, \eta)$ chart-- has proper length $D$. Thus $ (a_{2}/3)(\eta^\prime{}^3 - \eta_{\circ}^3) = D$. We will take the space-like separated points to lie on a $\eta=\eta_{\circ}$ surface such that the straight line connecting them --the geodesic within this surface-- has length $D$. Thus we will consider points $(\vx, \eta_{\circ})$ and $(\vxp, \eta_{\circ})$ such that $(a_{2}\eta_{\circ}^2)\, |\vx - \vxp| \equiv (a_{2}\eta_{\circ}^2)\, r = D$. The question is: How are the correlations associated with these two pairs of points related? For space-like separated points we have:
\be \label{2point5} \langle \hphi(\vx,\,\etao)\, \hphi(\vxp,\, \etao) \rangle
= \f{\hbar}{4\pi^2}\, \Big[\f{1}{D^2}  + \, \f{1}{a_{2}^2 \etao^6} \big(1-\gamma - \ln \f{D}{a_{2}\etao^2 \ell}\big)\, \Big]\, .\ee
Note incidentally that, even away from the singularity, as $D\to \infty$ for a fixed $\etao$, the correlations do not decay as the distance $D$ between the points goes to $\infty$; in fact they diverge logarithmically due to the infrared behavior we found in Section \ref{s4.1.1}. (Even if one were to adjust the infrared cut-off $\ell$ so that the logarithmic term is made to vanish for a given $\etao$, the correlations would approach a non-zero constant at $\eta=\etao$.) For time-like separations, we have
\be \label{2point6} \langle \hphi(\vx_\circ,\,\etao)\, \hphi(\vx_\circ,\, \etap) \rangle = - \f{\hbar}{4\pi^2} \f{1}{a_{2}^2\, \etao^2 \eta^{\prime\,2}} \Big[\f{1}{(\etao -\etap)^2} - \f{1}{\etao\etap}\, \big(1-\gamma - \ln \f{(\etap -\etao)}{\ell} \big)\, \Big] \ee
with $\etap = (\f{3D}{a_{2}}\, +\, \etao^3)^\f{1}{3}$. Let us now take the ratio of the two correlation functions and examine its behavior as we approach the singularity  by sending $\etao$ to zero. We obtain:
\be \label{ratio2} \lim_{\eta_{\circ} \to 0^+}\,\, \f{ \langle \hphi(\vx, \eta_{\circ})\, \hphi(\vxp, \eta_{\circ} )\rangle}{\langle \hphi(\vx_{\circ}, \eta_{\circ})\, \hphi(\vx_{\circ} \etap )\rangle\,}=\, -  \lim_{\eta_{\circ} \to 0^+}\,\,\,\, \f{3D}{a_2\etao^3}\, \ln \f{a_{2}\,\eta_o^2\, \ell}{D}\,\,{\Big[1-\gamma + \f{1}{3}\ln \f{a_{2}\ell^3}{2D}\Big]^{-1} } =\, \infty\, .\ee
Thus, for the dust filled FLRW universe, the correlations between fields at spatially separated points grow \emph{faster} than those between time-like separated points as one approaches the singularity, in striking contrast with the radiation filled FLRW model. 

What is the origin of this qualitative difference? It is clear from the calculation above that the difference can be traced directly to the last two terms in (\ref{2point4})-- which are absent in the radiation filled universe due to conformal covariance. Since conformal covariance is a peculiarity of the radiation dominated universe, extra terms are present in generic FLRW universe. Therefore, the qualitative behavior we found in this section is generic for FLRW models under consideration. Since higher correlations between fields at points $x$ and $\xp$ heuristically corresponds to a lower gradient between $x$ and $\xp$, for test quantum fields in a generic FLRW universe `time-derivatives dominate over space-derivatives near the space-like singularity'. In this sense, the generic behavior is along the lines of the classical BKL behavior in the vicinity of space-like singularities, although as emphasized in Section \ref{s3} there are also deep conceptual differences between the two. Conceptually, our quantum results are more closely related to the scaling of the `mutual information' in Gaussian states of test quantum fields that Bianchi and Satz found in the approach to the big bang singularity \cite{ebas}.

\subsection{Products of field operators}
\label{s5.3}

Finally, let us consider the distributions that result from the renormalized expectation values of products of field operators. These distributions have been discussed extensively in the older literature and the underlying procedure was summarized in Section \ref{s3.2.2}. Therefore we will only provide the main results, focusing on the effect of the big bang and big crunch singularity at $\eta=0$.
 
Recall that to calculate $\phisq$ one uses the point-splitting procedure that requires the 2-point bi-distribution $\biphi$ and the DeWitt-Schwinger counter-term. As in Section \ref{s3.2.2} let us follow the procedure of \cite{ori,valencia} and
split points along a translational Killing field tangential to the $\eta={\rm const}$ slices. Then we can use (\ref{2point5}) for the required point-split bi-distribution. The DeWitt-Schwinger counter-term is now given by
\ba G_{\rm DS}(x,\xp) &=&\frac{\hbar}{8\pi^2\sigma} - \f{\hbar}{48\pi^2} (\gamma + \f{1}{2} \ln \f{\mu^2\sigma}{2}) + \frac{\hbar}{96\pi^2\sigma}R_{ab}\sigma^a\sigma^b \nonumber\\
&=& \f{\hbar}{4\pi^2} \Big[\f{1}{D^2} + \f{1}{a_{2}^2 \eta^6}\, \Big(\f{1}{6} - \gamma - \ln \f{\mu\,D}{2} \Big)\, + \mathcal{O}(r)\Big]\ea
where as before $2\sigma$ is the signed square of the geodesic distance between $x$ and $\xp$,\, $\sigma_a = \nabla_a \sigma(x,\,\xp)$,\, $r = |\vx -\vxp|$, and $\mu$ is the DeWitt-Schwinger ultraviolet cut-off. Therefore, $\phisq$ is given by
\ba\label{SET2} \phisq &=& \lim_{x \to \xp}\, \big[\langle\phi(x)\phi(\xp)\rangle\, - \, G_{\rm DS}(x,\, \xp) \big] \nonumber\\
&=& \f{\hbar R}{288\pi^2}\, \Big(5 -2\,\ln\, \f{2R}{3a_{2}\ell^3 \mu^3}
\Big)\,.\ea
Regarded as a function, $\phisq$ diverges at the singularity $\eta=0$ because the scalar curvature is given by $R = {12}/(a_{2}^2\, \eta^6)$. However, it is a well-defined tempered distribution on $(\Mo, g_{ab})$ because the volume element of $g_{ab}$ is given by $\rmd^4 V = a_{2}^4\, \eta^8\, \rmd^4 x$: Given a test field $f(x)$, we have
\be \phisq\, :\,\, f(x) \rightarrow  \f{\hbar a_2^2}{24\pi^2}\, \int_{\Mo} \rmd^4 x\,  \eta^2\,\big( 5 - 2 \ln 8 +6 \ln a_2\ell\mu\, \eta^2 \big) f(x) \, .\ee
Thus, the action of $\phisq$ on a test function $f(x)$ is the same as that of a $C^1$ function of $\eta$ on that $f(x)$.
\smallskip

Next, let us consider $\stress$. The renormalized stress energy tensor for a massless Klein-Gordon field in a FLRW space-time with scale factor $a(\eta) = a_{\beta}\, \eta^{\beta}$ was calculated by Bunch and Davies%
\footnote{Note however that they use a metric with signature $(+,-,-,-)$ and their convention for the Riemann tensor is $2\nabla_{[a} \nabla_{b]} K_c = -2 R_{abc}{}^d K_d$. Therefore, their derivative operator and scalar curvature is the same as ours while their metric, Riemann and Ricci tensors and $\Box$ carry a negative sign relative to ours. The expression (\ref{SET2}) contains constants that are expressed as digamma functions of $\beta$ that determines our FLRW model. There are poles in the digamma functions for certain values of the argument but the three terms in the first square bracket can be regularized by introducing an infrared cut-off. See, e.g., \cite{bd}.}
in \cite{bd2}:
%
%
\ba \label{SET2} \stress &=& \,\frac{\hbar}{69120\pi^2}\,\big[-168\nabla_a\nabla_bR
+ 288\,\Box R\,g_{ab} - 24\,R_{ac}{R^c}_b + 12\,R^{cd}R_{cd}\, g_{ab}\nonumber\\
&+& \, 64R\,R_{ab} \big]
- \frac{\hbar}{1152\pi^2}\, \big[\,({21}/{20}) +  \f{1}{\beta(\beta - 1)}\,\big] R^2\, g_{ab}\, \nonumber \\
&-& \frac{\hbar}{1152\pi^2}\, \big[\ln \big({|R|}/{\mu^2}\big)\, + \psi(1+\beta) +\psi (2-\beta)\big]\, {}^{(1)}H_{ab}
\ea
where $\psi$ is the digamma function of its argument (and thus $\eta$ independent)
%
%
and $H_{ab}$ is given by:
\be {}^{(1)}H_{ab} = 2 \nabla_a \nabla_b R\,\, - 2\,\Box R\, g_{ab} - 2\,\big(R R_{ab}\, - \,(R^2/4)\, g_{ab} \big)\,. 
\ee
Because of the FLRW symmetries, the right side of $\stress$ has only two independent components whence we can again write it as 
\be  \stress = T_1(\eta) \nabla_a\eta\, \nabla_b \eta\, +\, T_2(\eta)\, \go_{ab}\,  \ee
following Section \ref{s3.2.2}. Since $\nabla_a \eta$ and $\go_{ab}$ are smooth, whether $\stress$ is a tempered distribution depends on the coefficients $T_1(\eta)$ and $T_2(\eta)$. Given a test field $f^{ab}(x) = f_1(x) \eta^a\eta^b\, +\, f_2(x) \go^{ab}$, with $f_1(x)$ and $f_2(x)$ in $\S$, we have:
\be \label{SET3} \stress\, :\,\, f^{ab}\,\, \rightarrow\,\, \int_{\Mo} \rmd^4 \Vo (a_2^4\, \eta^8)\, (T_1 f_1 + 4 T_2 f_2)\, .\ee
For a dust-filled universe, curvature tensor is such that each term in the first two lines of (\ref{SET2}) diverges (at most) as $1/\eta^{8}$ at $\eta=0$. The most divergent piece in ${}^{(1)}H_{ab}$ also goes as $1/\eta^8$ while the scalar curvature $R$ goes as $1/\eta^6$. Therefore, the most divergent term in $T_1$ and $T_2$ goes as $({1}/{\eta^8}) \ln |\eta|$. Because of the $\eta^8$ term coming from the volume element, then, the right side of (\ref{SET3}) is given by the distributional action of a locally integrable function, $(\ln |\eta| +$ functions that are regular at $\eta=0)$, on test functions $f_1(x)$ and $f_2(x)$. Therefore we conclude that $\stress$ is well-defined as a tempered distribution. Interestingly, whereas $\stress$ for the radiation-filled universe features distributions that go as $1/\eta^2$, for the dust filled universe (and for $\beta > 2$) $\stress$ is a tamer distribution.


\section{Discussion}
\label{s6}

Within classical theories of gravity, space-like singularities are generally considered as absolute barriers that mark the beginning or end of space-time. 
In particular, geodesics followed by test particles meet their end, and the tidal forces between them grow unboundedly as they approach curvature singularities. One can also use test classical fields to probe the space-time geometry in the vicinity of these singularities. Again, these probes strongly sense the presence of the singularity and diverge there (unless the space-time is conformally flat and equations of motion satisfied by these fields are conformally invariant). 
Would quantum probes also sense this singularity? or, would the quintessential quantum fuzziness somehow soften its potency, allowing them to be regular across the singularity? These questions has been raised in the literature by several authors over the past three decades. In this paper we investigated the issue in considerable detail for the big bang and big crunch singularities of FLRW cosmologies. There are some key differences between our analysis and previous studies of effects of space-time singularities on quantum probes. For example, in contrast to the analysis of, e.g., \cite{hm,stz}, the singularity results from dynamical evolution, is space-like, and physically more interesting. Also, our probes are quantum fields rather than quantum particles. Similarly, in contrast to the investigation of the Schwarzschild singularity using quantum fields as probes of \cite{hs}, we pay due attention to mathematical issues that arise due to the presence of an infinite number of degrees of freedom of quantum fields, without taking recourse to, e.g., formal integrations on infinite dimensional spaces. 

We found that, although classical fields $\phi(x)$ and the mode functions $e(k, \eta)$ that are commonly used in the analysis of the quantum fields $\hphi(x)$  do diverge at the big-bang, the quantum field $\hphi(x)$ is \emph{well behaved across the big bang and the big crunch as an operator valued tempered distribution.} They also satisfy (the expected commutation relations and) the field equation $\Box\, \hphi(x) =0$ in the distributional sense on the extended space-time.%
\footnote{As we pointed out in sections \ref{s3} - \ref{s5}, the classical solutions $\phi(x)$ and mode functions $e(k, \eta)$ are also well defined distributions. We could have worked with rescaled fields $\underline{\phi}(x) := \sqrt{g}\phi(x) \equiv a^4(\eta)\phi(x)$. The rescaled field would have been well-defined everywhere on the extended manifold $\Mo$. But it would carry a density of weight $1$ making the discussion of the phase space $\ps$, and of the K\"ahler structure thereon, rather awkward.}
Note that even in Minkowski space-time $(\Mo, \go_{ab})$,\, $\hphio(x)$ is not an operator but an operator valued tempered distribution. It has to be integrated against a test function $f(x)$ in the Schwartz space $\S$ to obtain an operator $\hphio(f) = \int_{\Mo} \rmd^4 x\, \hphio(x) f(x)$. Similarly, the expectation value $\langle \hphio(x) \hphio(\xp) \rangle_{{}_\circ}$ is a tempered bi-distribution. We showed in detail that in the radiation and dust-filled 
FLRW universes $(\Mo, g_{ab})$, operators $\hphi(f) = \int_{\Mo} \rmd^4 V \hphio(x) f(x)$ are well-behaved even when the test fields $f(x)\in \S$ have support that includes the singularity. In essence, this is because, although the mode functions $e(k, \eta)$ do diverge at the big-bang $\eta=0$, their divergence is (more than) compensated by the fact that the volume element\, $\rmd^4 V = a^4(\eta) \rmd^4 x$\, on $(\Mo, g_{ab})$ goes to zero at $\eta=0$ sufficiently fast. Similarly, the classical solutions $F(x)$ --that define elements of the 1-particle Hilbert space $\H$-- diverge at $\eta=0$, if regarded as functions. However, they correspond to well-defined elements of $\H$ since their norm with respect to Hermitian inner product on $\H$ is finite, again because the volume element shrinks at the big bang. This is analogous to the fact that, while the quantum mechanical  wave function $\psi(\vx) := (1/r) e^{-\alpha\, r}$ (with $\alpha >0$) diverges at the origin, it is an admissible quantum state because its norm in $L^2(\mathbb{R}^3)$ is finite.

In the dust-filled universe (and more generally for $a(\eta) = a_\beta\, \eta^\beta$, with $\beta \ge 2$), there is a well-known infrared issue \cite{fp} that occurs away from the singularity; it is mild in the sense that the action of $\hphi(x)$ is already well defined on an infinite dimensional subspace $\S_1$ of $\S$ with co-dimension $1$ and we need an infrared cutoff only to extend this action to all of $\S$. Once this issue is handled away from the singularity, the presence of the singularity does not introduce any further complication (because the potential problems due to the presence of singularity refer to the ultraviolet behavior of the field rather than infrared). 

Similarly, we found that the 2-point bi-distribution $\biphi$ is well-behaved in spite of the singularity. When the points $x$ and $\xp$ are space-like or time-like separated, the bi-distribution provides correlation functions. We found that, as one approaches the singularity, there is an interesting asymmetry between the correlation functions associated with space-like and time-like separated points, similar to the asymmetry one finds between spatial and temporal derivatives of geometric fields in the BKL behavior. However, conceptually, the two features are quite different. Our calculations refer to the behavior of test quantum fields on a given FLRW space-time while the BKL behavior refers to the Einstein dynamics of the gravitational field itself.
  
We also investigated the renormalized expectation values $\phisq$ and $\stress$ of products of OVDs in these space-times. We found that they continue to be well defined tempered distribution across the big bang and big crunch. In the radiation-filled FLRW universe the scalar curvature vanishes, whence solutions $\hphi(x)$\,  to\, $\Box\, \hphi(x) =0$\, on $(\Mo, g_{ab})$ are related to the solutions $\hphio(x)$ of\, $\Boxo\, \hphio(x) = 0$\, in Minkowski space-time $(\Mo, \go_{ab})$ simply by $\hphi(x) = \hphio(x)/a(\eta)$. One may therefore be tempted to think that the tameness of observables associated with $\hphi(x)$ in this space-time is not too surprising. Indeed, this point served as a key motivation for us to investigate the quantum field $\hphi(x)$ in the dust filled universe where $\hphi(x)$ is not related  to $\hphio(x)$ in Minkowski space in any simple way. Not only are $\hphi(x)$ and $\phisq$ also tame in the dust-filled universe but, surprisingly, $\stress$ is in fact better behaved than in the radiation-filled universe: As a distribution, it features only locally integrable functions on $\Mo$ rather than `genuine' distributions such as $\underline{\eta}^{-2}$.

We restricted the detailed analysis to the radiation and dust filled universes because in these cases the mode functions have a simple closed form, making the analysis technically simpler and allowing us to obtain explicit expressions of $\hphi(x), \, \biphi,\, \phisq$ and $\stress$. These in turn enabled us to focus on the conceptual issues that are central to the main question. However, as we briefly discuss in Appendix \ref{a2}, the tameness of the space-like singularity extends to scalar fields on more general space-times, as well as to higher spin fields. 

In this analysis, to ask whether observables are well-behaved across the singularity, we needed to extend physical space-time $(M,g_{ab})$\, --with $\eta >0$ and a singularity at $\eta = 0$-- \,to the extended space-time $(\Mo, g_{ab})$, with $\eta \in (-\infty, \infty)$. Conformal flatness of FLRW space-times provides a natural extension $\Mo$ of $M$. What would happen in non-conformally flat space-times such as, for example, the Bianchi models or the Schwarzschild space-time? It turns out that there is a generalization of the standard Hamiltonian formulation of general relativity based on fields with only `internal $SO(3)$ indices' \cite{ahs1,ahs2}, that is well-suited to provide the required extension. When the 3-metric is invertible, this formulation is equivalent to the standard Arnowitt, Deser, Misner (ADM) framework. However, even if the covariant 3-metric becomes degenerate (causing  its curvature to diverge) the equations satisfied by the fields with only `internal indices' do not break down. Thus, the new framework provides a generalization of Einstein's equations in the ADM form that permits one to extend space-times across certain space-like singularities. In the FLRW space-times, this extension coincides with the `obvious' one. However, it also enables one to extend certain non conformally-flat space-times such as Kasner and Kantowski-Sachs universes \cite{sloan,mercatti,nv}. Therefore one can ask if test quantum fields remain well behaved on these extensions. 

The Kantowski-Sachs space-time is especially interesting because it is isometric to  the Schwarzschild-interior (i.e., the portion of the Schwarzschild space-time inside the horizon). We have investigated this case. All 4 Killing fields are now space-like and tangential to the homogeneous 3-manifolds $r = {\rm const}$ (with $r$ playing the role of time). One can introduce a basis of solutions to \, $\Box\, \phi =0$\, of the form $e_{\ell}(k, r)\, e^{ikt}\, Y_{\ell, m}(\theta,\phi)$, where $k \in R$ and $e_{\ell}(k, r)$ are now the mode functions (analogs of $e(k, \eta)$ in FLRW space-times). For each choice of $k$ and $\ell$,\, $e_{\ell} (k, r)$ is subject to a second order linear differential equation in $r$. We are able to write down an analytical expression of the two independent solutions as an infinite convergent series. One of the solutions is regular across the singularity while the other diverges, but only logarithmically in $r$. Since the volume element vanishes as $r^2$ at $r=0$, the basis functions are well-defined tempered distributions as in the FLRW case. This enabled us to introduce a covariant phase space $\ps$ with a well-defined symplectic structure, just as in the FLRW space-times, and we constructed the corresponding operator algebra, generated by $\hPhi(F)$ where  $F(x) \in \ps$. We could just choose a complex structure $J$ on $\ps$ and construct the corresponding Fock representation on which $\h\Phi(F)$ are represented, as usual, by sums of creation and annihilation operators.  In this sense, the situation would be the same as in FLRW space-times; the Schwarzschild singularity also appears to be rather tame when probed with quantum fields $\h\Phi(F)$. It should be possible to write down an operator-valued distribution $\hphi(x)$ so that the smeared $\hphi(f)$ are also well-defined operators on the Fock space. Detailed considerations may again show that we have to restrict the test functions $f$ to a subspace of the Schwartz space $\S$ of finite co-dimension and then extend the action to full $\S$ with additional, well-motivated inputs. The key open issue is the following: so far we do not have a principle or a procedure (analogous to that of Section \ref{s4.1.2}) to select a \emph{preferred} complex structure. Physically, one would like to introduce that complex structure which corresponds to the Unruh vacuum. However, singling out this complex structure is rather subtle because the Schwarzschild horizon is not a part of the Kantowski-Sachs space-time. Once a physically well-motivated complex structure is found, one would be able to analyze in detail the behavior of the corresponding $\biphi$,\, $\phisq$ and $\stress$ across the singularity. For more general black hole singularities, one would have to consider evolution across singular Cauchy horizons, which are null rather than space-like. However, there are already discussions of the required space-time extensions in the literature (see, e.g., \cite{ori2,md}) and it would be very interesting to investigate the evolution of quantum fields in these extensions, across the Cauchy horizon.

To summarize then, the answer to the question we set out to investigate is the following: the most interesting space-like singularities appear to be much tamer when probed with quantum fields than they are when probed using classical particles or fields. Tidal forces between test particles and observables constructed from classical fields do diverge at the singularity. But these need not be the appropriate tools. It is generally believed that quantum fields would be physically more appropriate.

However, we would also like to emphasize that this analysis has been carried out in a hybrid framework where the geometry is treated classically and probes quantum mechanically. A full, self-consistent theory has to treat both quantum mechanically, and allow them to interact in a consistent manner. This is the burden of a satisfactory theory of quantum gravity. Results such as ours can serve as guidelines in that they help sharpen the questions of where the current incompleteness lies. It is not that we need quantum gravity because the quantum theory of fields in curved space-times drastically fails at the physically most interesting singularities. Rather, this theory is physically inappropriate in the deep Planck regime because it takes into account only the quantum nature of matter and not of geometry. Therefore, it is of great physical interest to know what in fact happens to quantum geometry in \emph{generic} situations in which the classical metric becomes singular in general relativity. Is quantum geometry nonetheless well-defined in a distributional sense? There are hints from several different approaches that quantum geometry is supported on 2 rather than 4 space-time dimensions at the micro-level \cite{carlip}. In particular, loop quantum gravity and spin foams provide a detailed realization of such a quantum Riemannian geometry at the Planck scale (see, e.g., \cite{30years1, 30years2}). Our results may help bridge the gap between these distributional quantum geometries  and quantum field theory in curved space-times.

\section*{Acknowledgments}
This work was supported by the NSF grants PHY-1505411 and PHY-1806356, the Eberly Chair funds of Penn State, and the Alexander von Humboldt Foundation.

\begin{appendix}
\section{Distributions, the Cauchy principal value, and anti-derivatives}
\label{a1}
For convenience of the reader, in this Appendix we will recall some standard results on tempered distributions from \cite{schwartz,gs,hormander,loja} which cannot be easily found in the physics literature.

The Schwarz space on $\mathbb{R}^n$ consists of all $C^\infty$ functions $f$ such that  $x^k D^m f$ is bounded on $\mathbb{R}^n$ for every integer $k$ and $m$. Here $x^k$  denotes products of order $k$ of the Cartesian coordinates on $\mathbb{R}^n$ and $D^m$ denotes any combination of derivatives of order $m$ w.r.t. these coordinates. $\S$ is endowed the a family of semi-norms $P_{k,m} = {\rm sup}_{x\in \mathbb{R}^n}\, \mid x^k\, D^m f(x)\mid $. This family induces a topology on $\S$. A tempered distribution $\phi$ on $\mathbb{R}^n$ is a continuous linear map from $\S$ to $\mathbb{C}$ in this topology.

A locally integrable function $e(x)$ on $\mathbb{R}^n$ defines a tempered distribution via: 
\be e(x):\,\, f(x) \to \int_{\mathbb{R}^n}\!\rmd^n x\,\, e(x)\, f(x)\quad {\rm for\,\, all}\,\,\,f(x) \in \S\, .\ee
In particular then, the function $e(x) = \ln |x|$ is a tempered distribution on $\mathbb{R}$. Derivatives of a given tempered distribution to arbitrary orders define other tempered distributions. These considerations provide tempered distributions $\ub{x}^{-m}$ (with $m$ a positive integer) via:
\be \label{xminusm} \ub{x}^{-m} = \f{{(-1)}^{m-1}}{(m-1)!}\,\f{\rmd^m \ln |x|}{\rmd x^m};\quad {\rm i.e.},\quad
\ub{x}^{-m}:\,\, f(x) \quad\rightarrow\quad  - \f{{1}}{(m-1)!}\, \int\!\rmd x\, \ln |x|\,\,\f{\rmd^m f}{\rmd x^m} \,\, . \ee
This definition of $\ub{x}^{-m}$ is completely analogous to the definition of the more familiar distribution $\delta^m(x)$ --the m-th derivative of the Dirac distribution-- which is is defined as the $m+1$th derivative of the locally integrable but non-differentiable step function. Finally, in practice, it is extremely useful that $\ub{x}^{-m}$ `interact' with the operation of taking derivatives and of multiplication by functions $x^n$ in the familiar way:
\be \label{properties1} \f{\rmd}{\rmd x}\,\ub{x}^{-m}\, =\, -m\, \ub{x}^{-m-1};\qquad \hbox{\rm and,\,\, if $m> 1$, then}\qquad x\, \ub{x}^{-m} = \ub{x}^{-m+1}\, . \ee
(For $m=1$, one has \, $x\, \ub{x}^{-1} =1$, i.e., $x\, \ub{x}^{-1}: f(x) \to \int f(x)\rmd x$.) These properties are useful in checking that distributions --such as the basis $e_\beta(k, \eta)$-- satisfy the desired equations such as (\ref{EOM4}).

Since the distribution defined by the function $1/\eta$ featured explicitly in some discussions in the main text, let us discuss the distribution $\ub{x}^{-1}$ further. 
In this case, the definition given above corresponds just to taking the Cauchy principal value of the integral:
\be \label{xminus1}\ub{x}^{-1}:\,\, f(x) \quad\rightarrow\quad  \lim_{\epsilon\to 0^+}\,\, \int_{\mathbb{R}\setminus [-\epsilon,\epsilon]}\!\!\rmd x \, \f{1}{x}\, f(x) \, , \ee
which can be re-expressed in forms that are often more directly useful in practice:
\be \label{xminus1a} \ub{x}^{-1}:\,\, f(x) \quad\rightarrow\quad \int_{0}^{\infty}\! \rmd x\,\, \f{1}{x}\, \big(f(x) - f(-x)\big)\, =\,\int_{-\infty}^{\infty}\!\rmd x\,\, \f{1}{x}\, \big(f(x) - f(0)\big)\, .\ee
(The first expression in (\ref{xminus1a}) brings out the fact that $\ub{x}^{-1}$ sends even test functions to zero,\, while the second makes the finiteness of the result of the action of the distribution manifest.) One can show directly that the right sides of (\ref{xminus1}) and (\ref{xminus1a}) are well-defined for all  $f\in \S$, and this linear mapping is continuous. Thus, the Cauchy principal value yields a tempered distribution which is the same as that defined in (\ref{xminusm}) for $m=1$. For higher values of $m$, the definition (\ref{xminusm}) of $\ub{x}^{-m}$ is a natural generalization of the Cauchy principal value in the sense that 
\be \label{xminuxm} \ub{x}^{-m}:\,\, f(x) \quad\rightarrow\quad \int_{-\infty}^{\infty} \!\rmd x\,\, \f{1}{x^m}\, \Big( f(x) - \sum_{n=0}^{m-1} \f{x^n}{n!}\, \f{\rmd^n{f}}{\rmd x^n}\Big)\, .   \ee
In the main text, $1/\eta$ and $1/\eta^2$ are the tempered distributions given by this construction.\medskip

This construction made use of the fact that the derivative of a tempered distribution is again a tempered distribution. It is natural to ask if one can invert the operation and take the \emph{anti-derivatives of tempered distributions}. The answer is in the affirmative and, as with integrals of functions, there is again a freedom to add an `integration constant'. Let us consider a distribution $e(x)$ on $\mathbb{R}$. We wish to define its anti-derivative $I(x)$ such that ${\rmd I}/{\rmd x} = e(x)$. It is natural to set $\int\! \rmd x \,I(x)\, ({\rmd f}/{\rmd x}) = - \int\! \rmd x\, e(x) f(x)$. This prescription determines the action of the desired distribution $I(x)$ on test fields $f_1(x)$ which can be written as $f_1 (x) = {\rmd f}/{\rmd x}$ for some $f\in S$. These $f_1(x)$ constitute a co-dimension 1 sub-space $\S_1$ of $\S$.%
\footnote{It is characterized by the fact that test functions $f_1(x)$ in $\S_1$ satisfy a single condition $\int \rmd x\, f_1(x) =0$. In terms of Fourier transforms, the space $\t{S}_1$ is the subspace on $\t{S}$ such that $\t{f}_1(k)\mid_{k=0} \,=0$.} 
To extend the action of $I(x)$ to full $\S$, one can proceed as follows \cite{schwartz}. Choose any $f_\circ(x) \in S$ that is \emph{not} in $\S_1$ and just \emph{define} the action of the extended $I(x)$ on this $f_\circ(x)$ to be a constant, say $c_\circ$. Now, any test function $f(x) \in \S$ can be uniquely written as $f(x) = b_\circ f_\circ(x) + f_1(x)$ where $b_\circ$ is a constant and $f_1\in \S_1$. Hence the action of $e(x)$ on any $f(x)$ is determined by linearity: $e(x):\, f\, \to\,\, b_\circ\,c_\circ \, -\, \int \rmd x\, I(x) f_1(x)$. One can check that the final answer is independent of the choice of the initial $f_\circ(x)$. Furthermore, the action can be shown to be continuous on $\S$, whence it provides an extension of the action of $I(x)$ from $\S_1$ to full $\S$ as a tempered distribution. $c_o$ is the `integration constant' that captures the freedom in the definition of the anti-derivative. In section \ref{s4.2.1} we encountered this freedom  through the `integration constant' $C_g$ in the expression (\ref{Istar}) of the anti-derivative $\t{I}^\star_g(k)$ of $\t{g}$ w.r.t. $k$, which was ultimately fixed through an infrared cutoff $\ell$.\\

\emph{Remark:} For completeness, we note that one can also promote the functions $x^{-m}$ to distributions through their \emph{algebraic} property that $x^{-m}$ is the inverse of $x^m$ (see e.g., \cite{schwartz,hormander,loja}). One is then led to the problem of defining the \emph{division} of a distribution by a function in $\S$: Is there a well-defined distribution $e(x)$ on $\mathbb{R}$ that corresponds to the division of the (trivial) distribution $e_\circ(x)=1$ by $x^m$? The answer is in the affirmative 
%
%
but the solution to the equation $x^m e(x) = 1$ is not unique. The general solution has a $m$ parameter  family of ambiguities, encoded in the constants $c_i$: 
\be \label{extension} e(x) = \ub{x}^{-m} + \sum_{i=0}^{m-1} c_i\, \delta^{(i)}(x) \ee
where $\delta^{(i)}(x)$ denotes the $i$th derivative of the Dirac distribution. The origin of this freedom is the following. It is clear that the obvious action $f \to 
\int\! \rmd x\, x^{-m}\, f(x)$\, is well-defined on the co-dimension $m$ subspace $\S_{m}$ of $\S$ consisting of test functions which vanish at $x=0$ together with their first $(m-1)$ derivatives, so that their Taylor expansion around $x=0$ starts with the term $ f_m x^m$. The problem is that of extending the action of the desired distribution to the full Schwartz space $\S$. This can be achieved by the obvious generalization of the procedure outlined above in our discussion of anti-derivatives and results in the freedom to choose the $c_i$ in (\ref{extension}). In the main body of the paper we chose the Cauchy principal value to define $1/\eta$ and $1/\eta^2$ as tempered distributions. This corresponds to setting $c_0 =0$ for $1/\eta$, which is motivated by the natural requirement that this distribution should send even test functions to zero (since $1/\eta$ changes sign under $\eta \to -\eta$). Similarly, with our choice, the distribution  $1/\eta^2$ sends odd test functions to zero. 

\section{More general models}
\label{a2}

In this Appendix we will provide a brief overview of directions in which the framework presented in the main body of the paper can be generalized. In each case, we will outline the main steps that are necessary for these extensions and indicate why we expect the `tameness' of cosmological singularities to persist.\\

$\bullet$\,\,\emph{General FLRW space-times:} In the main body of the paper we analyzed in detail the massless scalar fields in the radiation and dust filled FLRW universes. Let us now consider more general conformal factors $a(\eta) = a_{\beta}\, \eta^{\beta}$. Since our focus is on the big bang, we are led to assume $\beta >0$. For brevity, we will only sketch the line of reasoning and overlook technical subtleties  --e.g., associated with properties of Hankel functions for exceptional values of their order--  that would have required detours.

The basis functions selected by our complex structure\, $\Joub$\, of section \ref{s4.1.2} are given by the Hankel functions of second kind:
\be e_\beta(k, \eta) =   \f{(\pi \eta)^{\f{1}{2}}}{2a(\eta)}\,\, H^{(2)}_{\beta-\f{1}{2}} (k\eta) \ee
Let us first consider the case when $\beta$ is a positive integer. For $\beta \ge 2$, the expression of $e_\beta(k, \eta)$ has a finite number of terms that are singular at $\eta=0$, all of the type $\eta^{-m}$ where $m$ is a positive integer. Let us denote by $e_\beta (k, \ul\eta)$ the tempered distribution obtained by replacing each $\eta^{-m}$ in the singular terms by the tempered distribution $\ul{\eta}^{-m}$. This $e_\beta (k, \ul\eta)$ is a tempered distribution on the extended space-time $(\Mo, g_{ab})$ and satisfies the equation of motion (\ref{EOM2})  in virtue of properties (\ref{properties1}) (see footnote \ref{fn2}).

Therefore, as in sections \ref{s4.1} and \ref{s5.1}, we can introduce the phase space $\ps$ using solutions $F(x)$ of the type
\be \label{expansion4} F(x) = \int \f{\rmd^3 k}{(2\pi)^3}\, \Big(z(\vk)\,  e_\beta(k,\eta) + {z^\star}(-\vk)\, e_\beta^\star(k,\eta)\, \Big)\, e^{i\vk\cdot\vx} \ee 
where, as before, $z(\vk)$ belong to the Schwartz space $\t\S$ associated with the 3-dimensional momentum space. As in the main text, each $F(x)$ is a well defined distribution on the extended space-time $\Mo$. In fact, for any given $k$ the leading order divergence in the basis functions $e_\beta(k,\eta)$ goes as $\eta^{(1-2\beta)}$ as $\eta \to 0$ while the volume element shrinks as $\eta^{4\beta}$. Therefore, in the action of $F(x)$ on a test function $f(x)\in \S$, the integrand (w.r.t. the volume element $\rmd^4 x$ which is regular at $\eta=0$) is well defined on all of $\Mo$, vanishing as $\eta^{1+2\beta}$.

Starting with $\ps$, we can again construct the algebra $\A$ generated by the field operators $\hTheta (F)$ and represent it on the Fock space $\F$ selected by the basis functions $e_\beta(k, \eta)$. The question is whether there is a corresponding  operator valued distribution $\hphi(x)$. As in section \ref{s4.2.1}, this question can be rephrased as: Is there a 1-particle state $F(x)$ corresponding to every test function $f(x)$ in the Schwartz space? By repeating the analysis of that section step by step, one finds that there is an infrared issue for $\beta \ge 2$ because of the infrared behavior of the basis functions \cite{fp}. More precisely, the situation can be summarized as follows. The asymptotic expansion of the Hankel function near $k\eta=0$ implies that the leading order divergences of $e_\beta(k, \eta)$ w.r.t. $k$ goes as $k^{(\f{1}{2}-\beta)}$. Since, near $k=0$, the behavior of $z(k)$ constructed from a generic test function $f(x) \in \S$ is the same as that of $e_\beta(k, \eta)$, the integrand in $\int\! \rmd^3 k\, |z(\vk)|^2$ goes as $k^{1-2\beta}$. Because the volume element in the momentum space goes as $k^2$, the $z(k)$ constructed from a generic test function fails to be square integrable if $\beta\ge 2$. Thus, for $\beta \ge 2$, a test function $f(x)$ defines a solution $F(x)$ in the 1 particle Hilbert space only if it belongs to a subspace of $\S$ with finite co-dimension (in which  Fourier transforms of the test functions vanish as $k^n$ at $k=0$, with $n > (2\beta-4$)). One again needs a regulator $\ell$ to `tame' this infrared behavior. Once it is introduced, $\hphi(x)$ becomes a well-defined OVD on the Fock space $\F$. Therefore, the bi-distribution $\biphi$ is again well-defined. As in section \ref{s4}, the infrared regulator is already needed  away from the big bang singularity; the presence of the singularity does not make the behavior worse. To summarize, there is no ultraviolet difficulty in introducing the OVD $\hphi(x)$.  

To probe the behavior of observables $\phisq$ and $\stress$, let us begin by examining the $\eta$ dependence of the Ricci curvature. Since $g_{ab} = a_\beta^2 \,\eta^{2\beta}\, \go_{ab}$, one finds
\ba R_{ab}\, &=& \,\f{2\beta(\beta+1)}{\eta^2}\, \nabla_a\eta \nabla_b \eta \,+\, \f{\beta(2\beta -1)}{\eta^2}\, \go_{ab}; \quad {\rm and} \nonumber\\
R\, &=& \, \f{6\beta (\beta-1)}{a_\beta^2 \eta^{2+2\beta}}  \ea
Consequently, the leading order divergence in $\phisq$, regarded as a function, is given by \cite{bd2}
\be \phisq \sim\,\, R\, \ln R \,\, \sim \,\,\beta(\beta-1) \f{1}{\eta^{(2\beta+2)}}\, \ln |\eta|\, . \ee
Since the volume element goes as $\sim\, a_\beta^4\, \eta^{4\beta}$, the action of $\phisq$ on a test function $f(x)$ is given by the action of a locally integrable, continuous function that vanishes at $\eta=0$ like  $\beta(\beta-1)\,\eta^{(2\beta -2)}\ln |\eta|$. Next, let us consider the components of the renormalized stress energy tensor \cite{bd2}. Regarded as functions, their leading order divergence is given by:
\be \stress = T_1(x)\, \nabla_a\eta \nabla_b \eta \,+\, T_2(x) \go_{ab},\quad {\rm with}\quad T_1(x) \, \sim\, T_2(x) \, \sim \f{1}{\eta^{(2\beta+4)}}\, \ln (|\eta|) \ee
Since the volume element goes as $\sim\, a_\beta^4\, \eta^{4\beta}$, the action of $\stress$ on a test field $f_1(x) \eta^a\eta^b + f_2(x) g^{ab}$ is given by the distributional action of a locally integrable function $\eta^{2\beta-4} \, \ln |\eta|$ on the test functions $f_1(x)$ and $f_2(x)$. Thus, $\phisq$ and $\stress$ are well-defined distributions on $(\Mo, g_{ab})$.\smallskip

Finally, let us consider non-integer values of $\beta$. There is a tempered distribution $|\x|^{-\beta}$ (corresponding to functions $|\eta|^{-\beta}$) for each real number $\beta$ that is \emph{not} a positive integer. The previous strategy of taking derivatives of a known distribution does not work. But one can define  $|\x|^{-\beta}$ using the fact that $|\eta|^{-\beta}$ is a homogeneous function of degree ${-\beta}$, although the procedure to define these homogeneous distributions \cite{hormander} is more complicated. One notes that for ${\rm Re}\, \beta < 1$, the function $|\eta|^{-\beta}$ is locally integrable and therefore defines a tempered distribution. Thus we have a map from the `${\rm Re}\, \beta < 1$ part' of the complex $\beta$ plane to the space of tempered distribution. It admits a unique meromorphic extension in $\beta$, that provides a tempered distribution $|{\x}|^{-\beta}$ for non-integral values of $\beta$ (the extension has simple poles at positive integer values of $\beta$). For our purposes, details of this procedure are not necessary; it suffices to note that these tempered distributions exist and satisfy the analogs of (\ref{properties1}) 
\be \label{properties2} \f{\rmd}{\rmd \eta}\, |{\x}|^{-\beta}\, =\, -\beta\, |{\x}|^{-\beta -1}\qquad \hbox{\rm and} \qquad  |\eta||{\x}|^{-\beta}= |{\x}|^{-\beta+1}\, . \ee
Let us denote by $e_\beta (k, \x)$ the tempered distribution obtained by replacing each $|\eta|^{-\beta}$ in the singular terms by the tempered distribution $\x^{-\beta}$. This $e_\beta (k, \x)$ is a tempered distribution on the extended space-time $(\Mo, g_{ab})$ and satisfies the equation of motion (\ref{EOM2})  in virtue of properties (\ref{properties2}). Therefore, one can proceed as in the case when $\beta$ is an integer and construct the covariant phase space, the algebra $\A$, and its Fock representation. Discussion of the bi-distribution $\biphi$ and observables $\phisq$ and $\stress$ is completely analogous. These observables are again well defined on the extended space-time $(\Mo, g_{ab})$ in the distributional sense.\\   

$\bullet$\,\, \emph{Other scalar field equations:} In the main body we focused on the minimally coupled, massless scalar field for technical simplicity. The most straightforward extension will be to the conformally coupled massless scalar field. The required analysis would be parallel to that of the radiation-filled universe of Section \ref{s3}. In a general FLRW space-time, if $F(x)$ satisfies $(\Box - \f{R}{6}) F(x) =0$, then $\Fo := a(\eta)F$ would satisfy $\Boxo\, \Fo (x) =0$. Therefore, one can construct the phase space $\ps$ of interest simply by rescaling the fields $\Fo(x) \in \pso$ satisfying the wave equation in Minkowski space by $1/a(\eta)$. As in Section \ref{s3}, one can then introduce the operator algebra $\A$ generated by $\hPhi(F)$ and represent these operators on the Fock space $\Focko$ of Minkowski space-time, construct the OVD $\hphi(x)$ and the bi-distribution $\biphi$ simply by conformally rescaling their Minkowski space analogs. Again, because the volume element goes as $\rmd^4 V = a^4(\eta)\,\rmd \Vo$,\, $\biphi$ would be a well-defined bi-distribution on $(\Mo, g_{ab})$. One would then be able to use the expressions of $\phisq$ and $\stress$ from \cite{dfcb}. 

These considerations can be easily generalized to conformally coupled scalar fields on \emph{general conformally flat space-times}, where the conformal factor is allowed to have any (smooth) space-time dependence. Thus, let us consider the Minkowski space-time $(\Mo, \go_{ab})$ and introduce on it a conformally flat metric $g_{ab} = \Omega^2(x) \go_{ab}$ such that the conformal factor $\Omega(x)$ goes to zero continuously on a Cauchy surface of $(\Mo, \go_{ab})$ --say, the $\eta=0$ hyperplane-- but is smooth elsewhere. Then, for the conformally coupled wave equation the analysis would be very similar as in the above discussion since each solution $\Fo(x)$ to the wave equation in Minkowski space-time would again define a (distributional) solution $F(x) = \Omega^{-1}\Fo(x)$ on $(\Mo, g_{ab})$. Construction of the phase space $\ps$ and the operator algebra $\A$ will go through. Since $\ps$ would be naturally isomorphic to the phase space $\pso$ in Minkowski space-time, one can use the complex structure $\Jo$ on $\pso$ to induce a complex structure $J$ on $\ps$. Because of the $\Omega^4(x)$ factor in the volume element $\rmd^4 V$, the operator valued distribution $\hphi(x)$ would again be well-defined and have a natural action on the Minkowski Fock space $\Focko$. 

Finally, let us consider the Klein-Gordon equation with mass $m$. Then, the potential $V(\eta)$ of Eq. (\ref{EOM3}) in Section \ref{s4} acquires an extra term and becomes $\t{V}(\eta) = V(\eta) - m^2a^2(\eta) \equiv {\beta(\beta-1)}/{\eta^2} - m^2 a^2(\eta)$. This makes the explicit calculations difficult. However, since the extra mass dependent term vanishes at $\eta=0$ where $V(\eta)$ diverges, one would expect that the main results on the well-defined character of various distributions across $\eta=0$ would not be altered by additional mass contribution to this potential, although there would be significant differences at early and late times where the new term would dominate. In particular, our arguments of Section \ref{s4.1.2} for selecting the complex structure $\Joub$ will not go through and new input will be needed to select the complex structure. But this issue refers to the quantization of the scalar field already on $(M, g_{ab})$ --i.e. for $\eta >0$-- and is unrelated to the principal issue for this paper --the behavior of the quantum field across $\eta=0$ on $(\Mo, g_{ab})$. Once a complex structure is chosen, operators $\hPhi(F)$ would be well-defined also the resulting Fock space.\bigskip

$\bullet$\,\, \emph{Higher spins:}\, Finally, let us consider higher spins. Since the Maxwell equation is conformally invariant and the FLRW metrics are conformally flat, every solution $\Fo_{ab}$ to Maxwell's equations on $(\Mo,\, \go_{ab})$ is a solution also on the extended FLRW space-time $(\Mo,\, g_{ab})$. Therefore, the OVD $\h{\Fo}_{ab}$ in Minkowski space itself defines the required OVD on $(\Mo,\, g_{ab})$ and it is trivially well-defined across the $\eta=0$ surface. Next consider linearized source-free solutions to Einstein's equations on FLRW backgrounds. Recall that each of the two polarization modes satisfies the massless wave equation on these backgrounds. Therefore, analysis of the paper is directly applicable to spin 2 fields. 

\end{appendix}

\end{document}